\documentclass[preprint]{aastex}
\usepackage{color}
\usepackage{natbib,graphicx,latexsym}

\newcommand{\Hipparcos}{{\sl Hipparcos}}

\newcommand{\HST}{{\sl HST}}

\newcommand{\Msun}{\mbox{M$_{\sun}$}}

\newcommand{\Lsun}{\mbox{L$_{\sun}$}}
\newcommand{\Rsun}{\mbox{R$_{\sun}$}}
\newcommand{\Mjup}{\mbox{M$_{\rm Jup}$}}

\newcommand{\ROSAT}{{\sl ROSAT}}

\newcommand{\kms}{\hbox{km~s$^{-1}$}}

\newcommand{\Ks}{\mbox{$K_S$}}
\newcommand{\Kcont}{\mbox{$K_{cont}$}}
\newcommand{\degs}{\mbox{$^{\circ}$}}
\newcommand{\Lbol}{\mbox{$L_{\rm bol}$}}

\newcommand{\Teff}{\mbox{$T_{\rm eff}$}}
\newcommand{\logg}{\mbox{$\log(g)$}}

\newcommand{\twomassbin}{\hbox{2MASS~J1534$-$2952AB}}
\newcommand{\hdbin}{\hbox{HD~130948BC}}
\newcommand{\hdprim}{\hbox{HD~130948A}}
\newcommand{\hdage}{\hbox{0.79$^{+0.22}_{-0.15}$~Gyr}}
\newcommand{\orbit}{\hbox{\tt ORBIT}}

\slugcomment{ApJ, in press}

\begin{document}

\title{Dynamical Mass of the Substellar Benchmark Binary HD~130948BC
  \altaffilmark{1,2,3}}

\author{Trent J. Dupuy,\altaffilmark{4}
        Michael C. Liu,\altaffilmark{4,5}
        Michael J. Ireland\altaffilmark{6}}

      \altaffiltext{1}{Most of the data presented herein were obtained
        at the W.M. Keck Observatory, which is operated as a
        scientific partnership among the California Institute of
        Technology, the University of California, and the National
        Aeronautics and Space Administration. The Observatory was made
        possible by the generous financial support of the W.M. Keck
        Foundation.}

      \altaffiltext{2}{Based on observations made with the NASA/ESA
        {\sl Hubble Space Telescope}, obtained from the data archive
        at the Space Telescope Institute. STScI is operated by the
        association of Universities for Research in Astronomy,
        Inc. under the NASA contract NAS 5-26555.}

      \altaffiltext{3}{Based on observations obtained at the Gemini
        Observatory, which is operated by the Association of
        Universities for Research in Astronomy, Inc., under a
        cooperative agreement with the NSF on behalf of the Gemini
        partnership: the National Science Foundation (United States),
        the Science and Technology Facilities Council (United
        Kingdom), the National Research Council (Canada), CONICYT
        (Chile), the Australian Research Council (Australia),
        Minist\'{e}rio da Ci\^{e}ncia e Tecnologia (Brazil) and SECYT
        (Argentina).}

      \altaffiltext{4}{Institute for Astronomy, University of Hawai`i,
        2680 Woodlawn Drive, Honolulu, HI 96822;
        tdupuy@ifa.hawaii.edu}

      \altaffiltext{5}{Alfred P. Sloan Research Fellow}

      \altaffiltext{6}{School of Physics, University of Sydney, NSW
        2006, Australia}

\begin{abstract}

  We present Keck adaptive optics imaging of the L4+L4 binary \hdbin\
  along with archival \HST\ and Gemini-North observations, which
  together span $\approx$70\% of the binary's orbital period.  From
  the relative orbit, we determine a total dynamical mass of
  0.109$\pm$0.002~\Msun\ (114$\pm$2~\Mjup).  The flux ratio of \hdbin\
  is near unity, so both components are unambiguously substellar for
  any plausible mass ratio.  An independent constraint on the age of
  the system is available from the primary \hdprim\ (G2V,
  [M/H]~=~0.0).  The ensemble of available indicators suggests an age
  comparable to the Hyades, with the most precise age being \hdage\
  based on gyrochronology.  Therefore, \hdbin\ is now a unique
  benchmark among field L and T~dwarfs, with a well-determined mass,
  luminosity, and age.  We find that substellar theoretical models
  disagree with our observations.  (1) Both components of \hdbin\
  appear to be overluminous by a factor of $\approx$2--3$\times$
  compared to evolutionary models.  The age of the system would have
  to be notably younger than the gyro age to ameliorate the luminosity
  disagreement.  (2) Effective temperatures derived from evolutionary
  models for HD~130948B and C are inconsistent with temperatures
  determined from spectral synthesis for objects of similar spectral
  type.  Overall, regardless of the adopted age, evolutionary and
  atmospheric models give inconsistent results, which indicates
  systematic errors in at least one class of models, possibly both.
  The masses of \hdbin\ happen to be very near the theoretical mass
  limit for lithium burning, and thus measuring the differential
  lithium depletion between B and C will provide a uniquely
  discriminating test of theoretical models.  The potential
  underestimate of luminosities by evolutionary models would have
  wide-ranging implications; therefore, a more refined age estimate
  for \hdprim\ is critically needed.

\end{abstract}



\section{Introduction}

More than a decade after their discovery, brown dwarfs continue to
offer key insights into the astrophysics governing of some of the
lowest temperature products of star formation.  Brown dwarfs in the
field are particularly useful as probes of very cold atmospheres.  For
instance, the atmospheres of extrasolar planets are very difficult to
study directly due to their intrinsic faintness and proximity to very
bright stars.  However, brown dwarfs are typically found in relative
isolation, and their atmospheres are subject to the same processes
(e.g., dust formation and sedimentation, and non-equilibrium molecular
chemistry) that are at work in their much less massive plantary
counterparts.

Despite the broad relevance of brown dwarfs, their fundamental
properties remain poorly constrained by observations.  In particular,
very few direct mass measurements are available for brown dwarfs.  To
date, a total of six objects have been identified as unambiguously
substellar \citep[$M<0.072$~\Msun\ at solar
metallicity;][]{2000ARA&A..38..337C} via dynamical mass measurements
with precisions ranging from 6--9\%: both components of the T5.0+T5.5
binary 2MASS~J15344984$-$2952274AB \citep{2008arXiv0807.0238L}, both
components of the young M6.5+M6.5 eclipsing binary
2MASS~J05352184$-$0546085 in the Orion Nebula
\citep{2006Natur.440..311S}, and two tertiary components of
hierarchical triples in which the primaries are M stars, GJ~802B
\citep{gl802b-ireland} and Gl~569Bb \citep{2001ApJ...560..390L,
  2004astro.ph..7334O, 2006ApJ...644.1183S}.  Direct mass measurements
of brown dwarfs are critical for empirically constraining substellar
evolutionary models.  Since brown dwarfs have no sustainable source of
internal energy, they follow a mass--luminosity--age relation, rather
than the simpler mass--luminosity relation for main-sequence stars.
Thus, mass measurements alone cannot fully constrain theoretical
models, although mass and luminosity measurements of brown dwarfs in
coeval binary systems can offer stringent tests of theoretical models
\citep[e.g.,][]{2008arXiv0807.0238L}.  To fully constrain evolutionary
models, systems with independent measurements of the mass, age, and
\Lbol\ (or one of the much less observationally accessible quanities
\Teff\ or $R$) are required.  Such systems are quite rare, but they
represent the gold standard among ``benchmark'' brown dwarfs.

\citet{2002ApJ...567L.133P} discovered the L~dwarf binary \hdbin\ in a
hierarchical triple configuration with the young solar analog \hdprim\
(G2V) using the curvature adaptive optics (AO) system Hokupa`a on the
Gemini North Telescope on 2001 February 24 UT.  The L~dwarfs are
separated from each other by $\lesssim$0$\farcs$13, and they lie
2$\farcs$6 from the primary G star.  \hdbin\ has been the target of
AO-fed slit spectroscopy with NIRSPEC on the Keck II Telescope
(1.15--1.35~\micron) and IRCS on the Subaru Telescope
(1.5--1.8~\micron, 1.95--2.4~\micron).  \citet{2002ApJ...567L..59G}
used the latter spectra to determine the spectral types of the B and C
components, both L4$\pm$1, via spectral template matching.  These are
consistent with the less precise NIRSPEC $J$-band spectral types of
dL2$\pm$2, which are on the \citet{1999AJ....118.2466M} system, found
by \citet{2002ApJ...567L.133P} for both HD~130948B and C.

We present here a dynamical mass measurement for \hdbin\ based on Keck
natural guide star adaptive optics (NGS AO) imaging of \hdbin, as well
as an analysis of {\sl Hubble Space Telescope} (\HST) and Gemini
archival images.  In addition to an independent age estimate (\hdage,
see \S \ref{sec:age}), the primary star provides a wealth of
information about the system.  \citet{2005ApJS..159..141V} measured a
solar metallicity for \hdprim\ ([M/H]~=~0.00, [Fe/H]~=~0.05), which is
important since metallicity can play a significant role in shaping the
spectra of brown dwarfs
\citep[e.g.,][]{2005astro.ph..9066B,2006liu-hd3651b}.  Most
importantly, the distance to \hdprim\ has been measured very precisely
by \Hipparcos, with a revised parallax of 55.01$\pm$0.24~mas
\citep{2007hnrr.book.....V}, corresponding to a distance of
$d$~=~18.18$\pm$0.08~pc.  Thus, the distance is measured to an
exquisite precision of 0.44\%, which is invaluable since the error in
the dynamical mass scales as 3$\times$ the distance error (i.e., the
0.44\% error in distance translates into a 1.3\% error in mass).

HD~130948BC can thus serve as both an ``age benchmark'' and ``mass
benchmark'' system in studying brown dwarfs.  In the literature, the
term benchmark is often applied to any readily observable unique or
extreme objects, but here we specifically use the term to refer to
systems for which fundamental properties may be directly determined.
\citet{2006MNRAS.368.1281P} highlighted the value of systems where the
age and composition of substellar objects can be independently
determined, e.g., from a stellar or white dwarf companion, and
\citet{2008arXiv0807.0238L} described an equivalent use of systems
with dynamical mass measurements.  Essentially, since brown dwarfs
follow a mass--luminosity--age relation, the measurement of either
mass or age in addition to the measured luminosity allows any other
quantity to be fully specified using evolutionary models.  This
approach can be extremely useful, for example, by offering precise
determinations of \Teff\ and \logg, which can then be compared
directly to atmospheric models.  Of course, the measurement of mass,
age, {\em and} luminosity offers a direct test of evolutionary models,
which is possible with \hdbin.


\section{Observations \label{sec:obs}}

\subsection{\HST/ACS-HRC Coronagraph \label{sec:hst}}

We retrieved {\sl Hubble Space Telescope}\ (\HST) archival images of
\hdbin\ obtained with the ACS High Resolution Camera (HRC) coronagraph
(1$\farcs$8 spot) on 2002 September 6 and 2005 February 23 UT.  These
data were taken as part of engineering programs to test the stability
of the PSF of an occulted star between orbits (ENG/ACS-10445, PI Cox)
and to measure the coronagraphic PSF as a function of wavelength
(CAL/ACS-9668, PI Krist).  Fortunately, the scientifically interesting
bright star \hdprim\ was selected to perform these tests.  At both
epochs, \hdbin\ is tight enough that the PSFs of the two components
are significantly blended with each other.  Therefore, to determine
the relative astrometry we fit all six parameters ($x$, $y$, and flux
for both components) simultaneously with an iterative,
$\chi^2$-minimum-finding approach.  Similar to our previous work
\citep{2008arXiv0807.0238L}, we used TinyTim
\citep{1995ASPC...77..349K} to model the off-spot PSFs of the binary
components and the \texttt{amoeba} algorithm
\citep[e.g.,][]{1992nrca.book.....P} for minimum finding in the
six-dimensional parameter space.

\begin{figure} 
\centerline{\includegraphics[width=6.5in,angle=0]{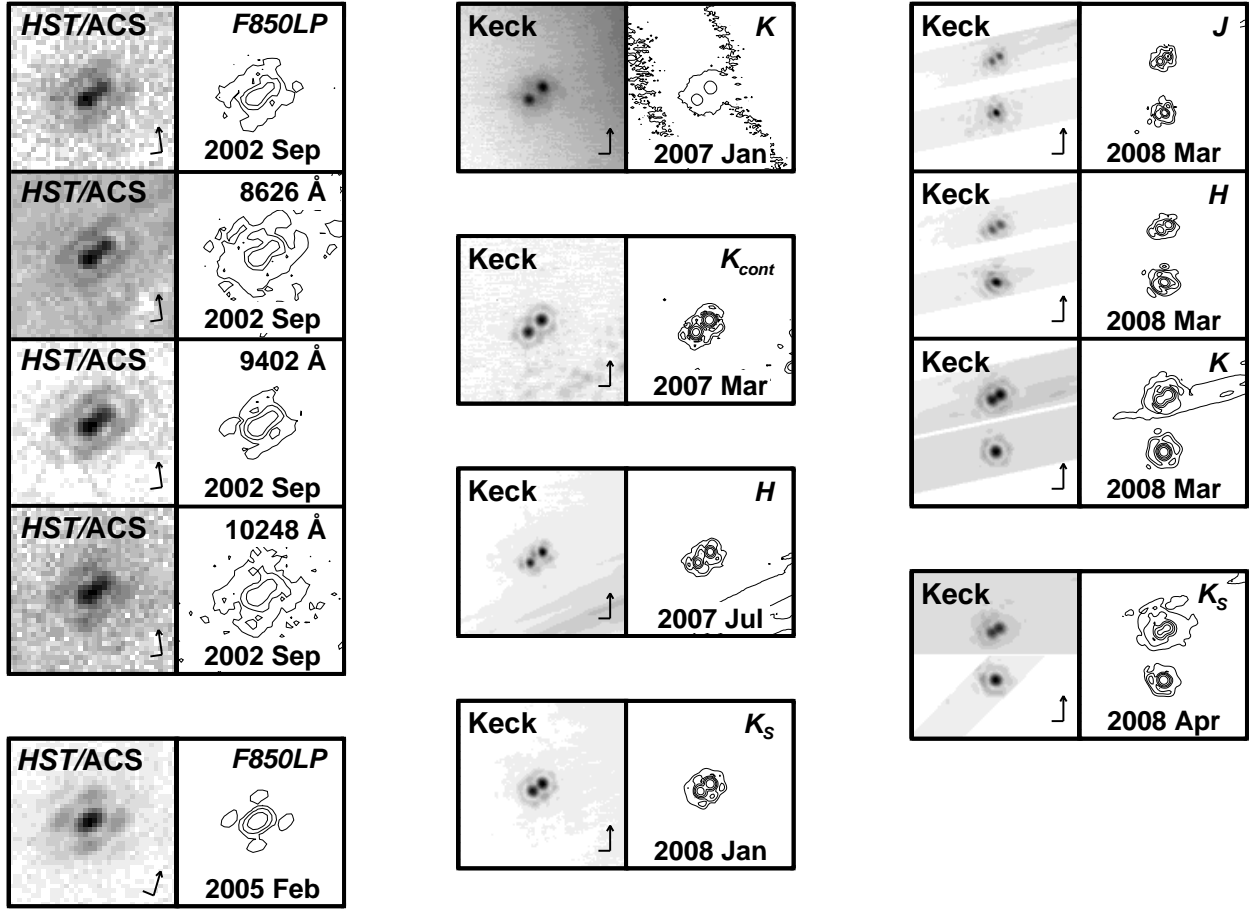}}

\caption{ \normalsize \HST\ and Keck images of \hdbin\ at all epochs
  and bandpasses, shown chronologically by column.  Unsaturated images
  of \hdprim\ are shown alongside the \hdbin\ images for the last two
  epochs.  For the most recent Keck epochs, interlaced short-exposure
  images of \hdprim\ are also shown.  For \HST\ images, the background
  light due to the occulted primary has been optimally subtracted as
  described in the text.  \HST\ images are in either $F850LP$ or the
  $FR914M$ ramp filter, which produces nearly-monochromatic images,
  and we have labeled the images with the wavelength (\texttt{LRFWAVE}
  header keyword).  We do not rotate the \HST\ data so that north is
  up in order to preserve the somewhat undersampled nature of the
  data.  Note that severe geometric distortion makes the cardinal
  directions in ACS-HRC images non-orthogonal.  In the Keck images,
  light from the PSF halo and/or the diffraction spikes of the primary
  are typically visible.  With the exception of the 2007 Jan epoch,
  during which seeing conditions were poor, the hexagonal Airy ring of
  the Keck PSF is visible in all the images. Both \HST\ and Keck
  images are shown on the same scale, 1$\farcs$0 on a side, with a
  square-root stretch for the grayscale images.  Contours are drawn at
  0.50, 0.25, 0.13, 0.06, and 0.03 of the peak pixel.  For lower $S/N$
  images (i.e., all \HST\ images and the 2007 Jan image) the two
  lowest contours are not drawn. \label{fig:images}}

\end{figure}

One challenge in obtaining precision astrometry for \hdbin\ is the
removal of background light from the primary, which is only 2$\farcs$6
away.  Though most of the light from \hdprim\ is occulted by the
1$\farcs$8 coronagraph spot, the remaining light reaching the detector
is highly structured and wavelength-dependent.  The most effective
technique for removal of the light from the occulted primary star is
to use an image taken at the same epoch but at a different telescope
roll angle.  This way, the background due to the bright primary
remains more or less unchanged while any other objects in the field
move to a different part of the image.  As part of the 2005
engineering tests, images were taken at two different roll angles, so
the background subtraction for these data is straightforward.
However, all of the engineering data from 2002 were taken at the same
roll angle, so data from a different epoch but the same filter must be
used for the background subtraction.  For the 2002 $F850LP$ data, we
were able to use the PSF of \hdprim\ itself because the 2005
engineering data were taken in this filter.  However, the remainder of
the 2002 data were taken with the linear ramp filter $FR914M$, and
\hdprim\ was never imaged in this filter on any other occasion.  Due
to the wavelength dependence of the background, we must use an
occulted star of similar spectral type to subtract the background.
Although very few $FR914M$ coronagraph data are available in the
archive, one star of identical spectral type, $\alpha$~Cen~A (G2V),
has been observed in $FR914M$.  The $\alpha$~Cen ramp filter data were
not taken at exactly the same nearly-monochromatic wavelengths as our
science images, so we used the two $\alpha$~Cen images which bracketed
our science data to create two different background-subtracted images
of \hdbin\ in $FR914M$.  Though the resulting background subtractions
were not very different, in the end, we only used the subtraction that
yielded a lower $\chi^2$ in the PSF-fitting of \hdbin.  When
performing any background subtraction, the images were first optimally
shifted and scaled to the nearest 1~pixel and 1\% in normalization by
minimizing the RMS of the subtraction residual of the central
200$\times$200-pixel region of the detector (this region excludes
\hdbin).  The final background-subtracted images of \hdbin\ are shown
in Figure~\ref{fig:images}.  The lack of visible structure in the
background demonstrates that the PSF from \hdprim\ was adequately
subtracted.

For PSF-fitting of the background-subtracted images, TinyTim PSF
models were generated to the specifications of the data.  PSFs were
created for detector locations within the nearest pixel of \hdbin\
(this is important because of the field dependence of the geometric
distortion that shapes the PSF).  We used the optical spectrum of the
L4 dwarf 2MASS~J0036+1821 from \citet{2001AJ....121.1710R} as the
spectral template for PSF generation.  We included different amounts
of telescope jitter (0 to 20 mas in 5 mas steps) and telescope defocus
($-$20 to $+$20 \micron\ in 4 \micron\ steps) to simulate the effect
of ``breathing'' on the PSF.\footnote{``Breathing'' is the term which
  has come to be used to describe the change in telescope focus due to
  thermal effects, especially within one pointing.  TinyTim
  parameterizes this as the offset between the secondary and primary
  mirrors of the telescope relative to the nominal in \micron.
  TinyTim also allows for different amounts of Gaussian telescope
  jitter due to guiding.}  We generated finely sampled PSFs at
5$\times$ the native pixel scale so that we could accurately
interpolate them to a fraction of a pixel.  We used the distortion
solution of \citet{ACS..ISR..2004-15} to correct the best-fit
positions for the severe geometric distortion of the ACS, and we used
their measured ACS pixel scale, which was derived by comparing
commanded (\texttt{POSTARG}) offsets of \HST\ in arcseconds to the
resulting pixel offsets.  They derived two such pixel scales for two
epochs of observations of 47~Tuc, and we adopt the mean and standard
deviation of these two: 28.273$\pm$0.006~mas/pix.

Rather than simply use the scatter of individual measurements to
estimate the astrometric uncertainty, we performed Monte Carlo
simulations in order to quantify potential systematic errors in our
PSF-fitting routine.  Typically, the most important source of
systematic error would be the imperfect modeling of the PSF, but for
\hdbin\ the imperfect subtraction of the background light due to
\hdprim\ may also be significant.  This is because the structured
background light will change even within a single orbit due to thermal
changes which affect the telescope optics (e.g., the well-known
``breathing'' phenomenon).  Also, the coronagraphic occulting spot
must be stowed when not in use, and each time it is deployed it will
be in a slightly different position, which will change the background
slightly, but noticeably.  It is easy to imagine how lumps in the
residual background structure resulting from an imperfect subtraction
could confuse measurement of both the astrometry and the flux ratio.
In Appendix~\ref{app:hst}, we describe in detail our Monte Carlo
simulations, from which we derived measurement offsets and
uncertainties in our PSF-fitting procedure for \HST/ACS images.  The
offsets ranged in amplitude from 0.1--0.8~mas in separation,
0.1--0.7\degs\ in PA, and 0.01--0.05~mag in flux ratio.  The offsets
are comparable in size to the uncertainties predicted by the Monte
Carlo simulations, which are also comparable in size to the RMS
scatter among the individual measurements.  The simulations showed
that systematic errors (i.e., imperfect PSF-modeling) dominate the
predicted uncertainties.  The best-fit positions and flux ratios from
both epochs, with offsets applied and uncertainties adopted from the
Monte Carlo simulations, are given in Table~\ref{tbl:hst}.

\subsection{Keck NGS AO \label{sec:keck}}

On 2007 January 26 UT, we began monitoring \hdbin\ using natural guide
star adaptive optics (NGS AO) at the Keck~II Telescope on Mauna Kea,
Hawaii.  The seeing conditions on that night were relatively poor;
however, at five subsequent epochs we obtained superior NGS AO imaging
data.  We used the facility IR camera NIRC2 with its narrow
field-of-view camera, which produces 10$\farcs$2$\times$10$\farcs$2
images.  The primary star \hdprim\
\citep[$R$~=~5.5~mag;][]{2003AJ....125..984M} located 2$\farcs$6 away
from \hdbin\ provided the reference for the AO correction.
Table~\ref{tbl:keck} summarizes our Keck NGS observations, and typical
images are shown in Figure~\ref{fig:images}.

At each epoch, \hdbin\ was imaged in one of the filters covering the
standard atmospheric windows from the Mauna Kea Observatories (MKO)
filter consortium \citep{mkofilters1,mkofilters2}.  We initially
obtained data in the $K$-band filter, though we subsequently took data
in $J$-, $H$-, $K_S$-, and $K_{cont}$-band
($\lambda_c$~=~2.271~\micron, $\Delta\lambda$~=~0.030~\micron).

On each observing run, we obtained dithered images, offsetting the
telescope by a few arcseconds between every 1--3 images.  There was no
need to exclude any of the images at any epoch on the basis of poor
image quality.  The images were reduced in a standard fashion. We
constructed flat fields from the differences of images of the
telescope dome interior with and without continuum lamp illumination.
Images were registered and stacked to form a final mosaic, though all
the results described here were based on analysis of the individual
images.

For the 2008 March and April epochs, we also obtained unsaturated
images of the primary \hdprim\ interlaced with deep exposures in which
it was saturated but \hdbin\ was measured at high $S/N$.  The minimum
integration time of NIRC2 is set by the sampling mode (e.g., single or
correlated double sampling) and how much of the array is read out, so
by using a very restricted subarray we achieved exposures of 4.372~ms
(512$\times$32) and 2.968~ms (384$\times$24).

We used the unsaturated images of the primary to estimate the Strehl
ratio and FWHM of images from the 2008 March and April epochs.  Since
no field stars were available to obtain a simultaneous measurement of
the Strehl ratio and FWHM in images from other epochs, we used images
of \hdbin\ itself.  The FWHM was determined by a Gaussian fit to the
core of the PSF of HD~130948B.  The quantities needed to calculate the
Strehl ratio were obtained as follows: (1) the peak flux of the
science PSF was determined by a Gaussian fit to the core of the PSF of
HD~130948B after removing contaminating flux from HD~130948C by
subtracting the science image from itself after being rotated by
180\degs\ about HD~130948C; (2) the total flux of HD~130948B was
measured using aperture photometry of \hdbin, then correcting for
binarity using the measured flux ratio for that epoch and filter; (3)
the peak-to-total flux ratio of the theoretically perfect PSF was
determined using the publicly available IDL routine
\texttt{NIRC2STREHL}\footnote{\url{http://www2.keck.hawaii.edu/optics/lgsao/software/nirc2strehl.pro}}
using the same aperture size as was used for the science image.  We
report the mean and standard deviation of the Strehl ratio and FWHM
measured for each set of dithered images at each epoch in
Table~\ref{tbl:keck}.

We computed the expected relative shift of the two components of
\hdbin\ at each epoch due to differential chromatic refraction (DCR).
Even though B and C have nearly identical spectral types based on
resolved spectroscopy, small color differences can change the extent
to which the atmosphere refracts their light, thus changing their
relative position as a function of airmass.  We used the prescription
of \citet{1992AJ....103..638M} for the effect of DCR, and we used the
prescription of \citet{1984A&A...138..275S} for the index of
refraction of dry air as a function of wavelength.  Thus, given the
effective wavelength of each component and details of the observations
(coordinates of \hdbin, time observed, and observatory latitude), we
calculated the expected shift due to DCR.  We computed the effective
wavelengths of B and C using template spectra of an L3 object
\citep[2MASS~J1146+2230AB;][]{2005ApJ...623.1115C} and an L5 object
\citep[SDSS~J0539-0059;][]{2005ApJ...623.1115C} to represent the
extremes of the possible spectral difference between the two
components of \hdbin\ (each is L4$\pm$1).  The resulting effective
wavelengths for $K$-band were 2.1976 \micron\ and 2.1939 \micron\ for
L3 and L5, respectively.  Even given this large allowance for spectral
differences between the two components of \hdbin, the DCR offset is
typically an order of magnitude smaller than the error at any given
epoch.  (The largest estimated offset is 0.5$\sigma$ for the 2008
April observations taken at airmass 1.6.)  Therefore, we are justified
in ignoring the effects of DCR in the relative astrometry.

To determine the relative positions and fluxes of \hdbin, we used a
simple analytic representation of the PSF to deblend the two
components.  The model was the sum of three elliptical Gaussians in
which each Gaussian component was allowed to have a different FWHM and
normalization, but all components had the same ellipticity and
semimajor axis PA.  In the vicinity of \hdbin\ there is a significant
contribution of both sky background and light from the bright primary
\hdprim.  Therefore, we also simultaneously fitted a sloped, flat
surface to account for the flux not due to \hdbin.  In all, we fitted
simultaneously for 16 parameters: 6 parameters for the positions and
fluxes of \hdbin, 7 parameters for the three-component elliptical
Gaussian model, and 3 parameters for background light due to \hdprim\
and the sky.  The best-fit parameters were found by a
Levenberg-Marquardt least-squares minimization in which all pixels
were weighted equally.  Our fitting procedure yielded a set of
measurements of the projected separation, PA, and flux ratio for
\hdbin.  We applied the distortion correction developed by B. Cameron
(priv. comm.) to the astrometry, which changed the results well below
the 1$\sigma$ level.

The internal scatter of the measurements at each epoch was very small,
but this does not include systematic errors due both to the imperfect
modeling of the PSF and to the temporally varying and spatially
structured background light.  To quantify these systematic errors, we
conducted extensive Monte Carlo simulations designed to replicate the
observations at each epoch.  These simulations are described in detail
in Appendix~\ref{app:keck}.  By comparing the input to fitted
parameters for 10$^3$ simulated images, we determined the offset and
uncertainty appropriate for each epoch.  The offsets ranged in
amplitude from 0.1--0.7~mas in separation, 0.03--0.20\degs\ in PA, and
0.001--0.1~mag in flux ratio.  We applied these offsets and although
they could be up to 1--3$\sigma$, their application only changed the
resulting total mass of \hdbin\ by only 0.2$\sigma$.  It also improved
the reduced $\chi^2$ of the best-fit orbit significantly, from a
reduced $\chi^2$ of 1.8 to a reduced $\chi^2$ of 1.1.  The
uncertainties predicted by the simulations were up to 3$\times$ larger
than the RMS scatter among individual measurements at each epoch.

For the 2008 March and April epochs, we used the interlaced
unsaturated images of the primary \hdprim\ in a different approach for
measuring the relative astrometry and flux ratio of \hdbin.  We
extracted 40$\times$80-pixel cut-outs of \hdprim\ from the short
exposures and \hdbin\ from the deeper exposures.  We then stitched
together these cut-outs, pairing an image of \hdbin\ with each of the
images of \hdprim\ taken immediately before and after it.  We employed
the StarFinder software package \citep{2000A&AS..147..335D} to
iteratively solve for the PSF, positions, and fluxes of the three
components in these combined images.  Unfortunately, the Keck AO
system did not keep \hdprim\ sufficiently fixed on the NIRC2 array (it
drifted up to $\sim$10~mas between deep exposures) to enable the
measurement of the positions of B and C relative to A from our data.
Since we used every available independent PSF contemporaneous with the
observations in measuring the system properties, we did not verify
this empirical PSF-fitting approach directly by testing StarFinder on
simulated binary images.  However, at the 2008 April epoch, we
obtained standard full-frame (1024$\times$1024) images, which we have
analyzed and simulated in the same fashion as data from the previous
epochs using our simple analytic model of the PSF.  The measured
relative astrometry and flux ratio are consistent using both
approaches.  In fact, the separation measured using our simple
analytic PSF model, which is more likely to be affected by systematic
errors, would be 3$\sigma$ discrepant with the StarFinder empirical
PSF results if the offset derived from simulated binary images was not
applied.  This suggests that our Monte Carlo simulations accurately
predict measurement offsets for the analytic PSF approach, and that
the StarFinder empirical PSF-fitting approach does not harbor any
significant systematic errors.

By interlacing short and deep exposures, we were able to measure the
photometric stability during each data set, thus enabling the
measurement of the absolute photometry of \hdbin\ by finding the
fluxes of B and C relative to A.  Photometry of \hdprim\ is contained
in the 2MASS Point Source Catalog \citep{2mass}, and we neglect the
small terms needed to convert the 2MASS photometry to the MKO system
for a G2 star ($\sim$0.003~mag).  Because \hdprim\ is very bright with
blue $JHK$ colors, it is unsaturated only in short 2MASS exposures
(51~ms) taken in $K$-band, thus, the quality of the 2MASS photometry
is best in $K$-band (0.02~mag uncertainty) and very poor in $J$- and
$H$-band (0.2~mag uncertainty).  To eliminate the need to rely on
2MASS $J$- and $H$-band photometry, we used $J-K$ and $H-K$ colors to
tie our $J$- and $H$-band photometry to the higher quality $K$-band
2MASS photometry.  We computed synthetic photometric colors of
$J-K$~=~0.339$\pm$0.010~mag and $H-K$~=~0.057$\pm$0.010~mag from a
low-resolution spectrum of \hdprim\ which we obtained using the IRTF
spectrograph SpeX \citep{1998SPIE.3354..468R} on
2008~May~16~UT\footnote{The spectrum of HD~76151 (G2V; Rayner,
  Cushing, \& Vacca 2007, in prep.) publicly available in the IRTF
  Spectral Library
  (\url{http://irtfweb.ifa.hawaii.edu/~spex/WebLibrary/}) yielded
  consistent synthetic colors.}.  The spectra were reduced using the
SpeXtool software package
\citep{2003PASP..115..389V,2004PASP..116..362C}.  Our adopted
uncertainties in the synthetic colors are the typical error in
determining the continuum slopes of FGK stars from $J$- to $K$-band
using SpeX (Rayner, Cushing, \& Vacca 2007, in prep.).  To determine
the uncertainty in the photometry of \hdbin, we added in quadrature:
(1) the error in the $K$-band 2MASS photometry of \hdprim; (2) the
error in the synthesized $J-K$ or $H-K$ color of \hdprim; (3) the
standard error of the mean relative flux measured in the
stitched-together images (i.e., the RMS scatter divided by $\sqrt{N}$,
where $N$ is the number of deep exposures given in
Table~\ref{tbl:keck}), which is dominated by the uncertainty in
deblending the B and C components; and (4) the RMS scatter in the
measured flux of \hdprim\ from our NIRC2 image sequence (a direct
measure of the photometric stability over the entire data set).  Our
photometry is consistent with that previously reported by
\citet{2002ApJ...567L.133P} but with much smaller errors in $J$- and
$K$-band.

In Table~\ref{tbl:keck}, we present the mean of the relative
astrometric and photometric measurements at each epoch as determined
from the two PSF-fitting procedures described above.  For cases in
which the analytic model of the PSF was used (prior to 2008 March),
offsets have been applied and uncertainties adopted from our Monte
Carlo simulations.  For the remaining cases, in which StarFinder
empirical PSF-fitting was used, the quoted uncertainty is the RMS
scatter of the measurements for individual images.  In
Table~\ref{tbl:phot}, we present our photometry for all three
components of HD~130948ABC.

\subsection{Gemini Hokupa`a AO \label{sec:gemini}}

As described by \citet{2002ApJ...567L.133P}, the discovery images of
\hdbin\ were obtained using the Hokupa`a curvature AO system on the
Gemini North Telescope, on Mauna Kea, Hawaii.  Hokupa`a observations
were carried out over a period of approximately 14 months beginning on
2001~February~24~UT.  We retrieved all available raw data from the
CADC Archive (GN-2001A-DD-2, GN-2001A-C-24, GN-2001B-DD-1,
GN-2002A-DD-1).  This included four epochs of dual-beam imaging, which
employed a Wollaston prism to simultaneously obtain orthogonally
polarized images, as well as two epochs of normal imaging in which the
Wollaston prism was not employed.  The Wollaston prism data were
originally used to search for circumstellar material around \hdprim\
using simultaneous difference imaging.  The nominal instrument
platescale, with or without the Wollaston prism, is
19.98$\pm$0.08~mas/pix.

We used the same PSF-fitting procedure as described for the Keck NGS
images on the Gemini data, which we registered and cosmic-ray
rejected.  As judged from the FWHM of the best-fit PSF model, the
image quality was best on 2001~February~24 and 2001~June~28~UT.
Simultaneous dual-beam images from 2001~February~24 yielded
inconsistent astrometry at the 2$\sigma$ level (0.7~mas) in separation
and the 6$\sigma$ level (1.2\degs) in PA, where these confidence
limits correspond to the internal scatter of the set of measurements
taken in each beam.  This inconsistency suggests that there are
significant systematic errors that affect the PSF, platescale, and/or
optical distortion of each orthogonally polarized beam differently.
As there is no way to effectively quantify these systematic errors, we
favor the use of the 2001~June~28 epoch images, which did not use the
Wollaston prism and are of comparable image quality
(FWHM~$\approx$~75~mas in $H$-band) to the 2001~February~24 data.  On
2001~June~28, the separation and PA measured by our PSF-fitting
procedure were 129.0$\pm$1.3~mas and 318.6$\pm$0.5\degs, respectively.
We refer to this as the ``measured'' astrometry hereinafter.  This
astrometry is consistent with the 134$\pm$5~mas separation and
317$\pm$1\degs\ PA reported by \citet{2002ApJ...567L.133P} as the
\textit{``average''} astrometry for the time period of
2001~February~24 to 2001~September~20~UT.

As an additional check, we also extracted the astrometry presented by
\citet{2003IAUS..211..265P} in their Figure 3. (The raw astrometry was
not published in that conference proceedings.)  They presented four
epochs spanning 2001~February~24 to 2002~April~23~UT, over which the
separation changes from 130.9 to 107.7~mas and the PA changes from
313.5 to 307.7\degs.  We refer to this as the ``extracted'' astrometry
hereinafter.  The separation range is consistent with what we measured
from the archival data, but the PA range is clearly inconsistent with
the PA we measured directly from the archival data (and the PA
reported by \citealp{2002ApJ...567L.133P}).  The $\approx$5\degs\
discrepancy in the PA could be explained if the orientation of the
detector on the sky was not correctly recorded in the header of the
archival data.  Indeed, this seems likely to be the case because all
archival data we retrieved record the same value for the orientation
of the detector (i.e., zero) even when the rotator had obviously been
changed by $\approx$90\degs\ for images taken at the same epoch.
However, this explanation fails to account for the inconsistency
between the PAs reported by
\citet{2002ApJ...567L.133P,2003IAUS..211..265P}.

Archival images of the binary Gl~569Bab taken on 2001~February~24~UT
using the same instrumental setup as the contemporaneous \hdbin\ data
(i.e., dual-beam Wollaston prism mode) offer strong evidence for the
sky orientation not being stored correctly in the header.  We measured
a PA of 70$\pm$3\degs\ for Gl~569Bab, whereas the ephemeris of
\citet{2006ApJ...644.1183S} predicts a PA of 314\degs\ at that epoch
(a 116\degs\ systematic offset!).  We also measured the separation of
Gl~569Bab to be 88$\pm$2~mas, which is discrepant with the ephemeris
prediction of 84~mas by 2$\sigma$.  This supports our suspicion of a
systematic error in the platescale of the instrument in Wollaston
prism mode that motivated us to use non-Wollaston images for our
``measured'' astrometry.

In the next section, we will consider both possible sets of Gemini
astrometry, ``extracted'' versus ``measured'', when determining the
orbit of \hdbin.  However, our default orbit solution uses only the
\HST\ and Keck astrometry.


\section{Dynamical Mass Determination \label{sec:orbit}}

\subsection{Orbit Fitting using Markov Chain Monte Carlo}

The orbit of \hdbin\ is quite well constrained as our observations
cover $\sim$70\% of the orbital period.  However, in order to search
for the influence of parameter degeneracies in our orbit fit and
determine robust confidence limits on the orbital parameters, we used
a Markov Chain Monte Carlo (MCMC) technique
\citep[e.g.,][]{bremaud99:markov_chain} for orbit fitting, in addition
to a gradient descent technique.  In short, the MCMC method constructs
a series of steps through the model parameter space such that the
resulting set of values (the ``chain'') is asymptotically equivalent
to the posterior probability distribution of the parameters being
sought.  The code that performed the MCMC fit is described in detail
in the study of \twomassbin\ by \citet{2008arXiv0807.0238L}.  Chains
all had lengths of 2$\times$10$^8$, and the correlation length of our
most correlated chain, as defined by \citet{2004PhRvD..69j3501T}, was
4.2$\times$10$^4$ for the orbital period, with equal or smaller
correlation lengths for other orbital parameters.  This gives an
effective length of the chain of 6.5$\times$10$^3$, which in turn
gives statistical uncertainties in the parameter errors of about
$1/\sqrt{6.5\times10^3}$~=~1.2\%, i.e., negligible.  We used uniform
priors in period ($P$), semimajor axis ($a$), PA of the ascending node
($\Omega$), argument of periastron ($\omega$), and time of periastron
passage ($T_0$).  We used a prior in inclination proportional to
$\sin(i)$ (i.e., random orbital orientation) and an eccentricity prior
of $f(e)$~=~2$e$ \citep[e.g., see][]{1991A&A...248..485D}.

As an independent verification of our MCMC results, we also fit the
orbit of \hdbin\ using the linearized least-squares routine \orbit\
\citep[described in][]{1999A&A...351..619F}.  We give the resulting
orbital parameters and their linearized uncertainties in
Table~\ref{tbl:orbit}.  All of the orbital parameters are consistent
between the \orbit\ and MCMC results.  In fact, they are all
consistent to much better than 1$\sigma$, which is expected since both
orbit-fitters should find the same $\chi^2$ minimum in parameter space
(i.e., the two methods are applied to the same dataset).  The reduced
$\chi^2$ of the \orbit\ solution was 1.14 (identical to the MCMC
solution), and the total mass was 0.108$\pm$0.002 \Msun, which
includes the error in the parallax and is also consistent with the
MCMC-derived mass.

\subsection{Fitting Results}

\begin{figure}
\vskip -1in
\centerline{\includegraphics[height=7.5in,angle=90]{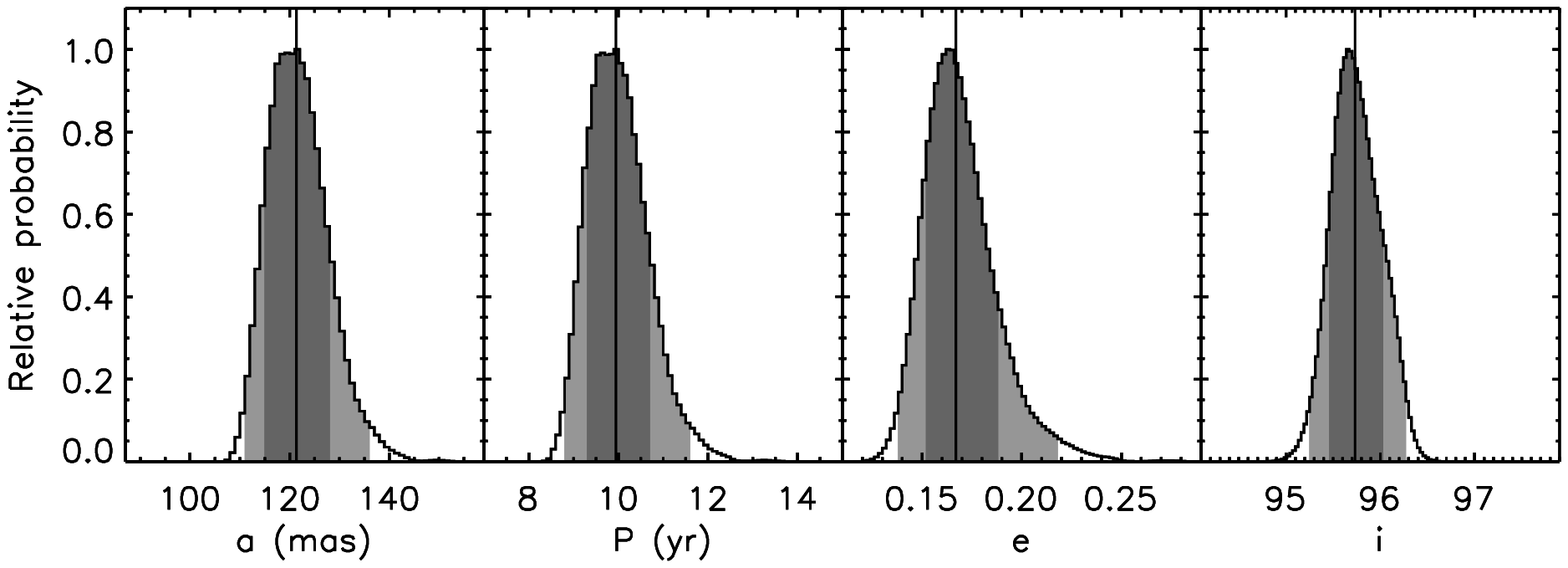}}
\centerline{\includegraphics[height=7.5in,angle=90]{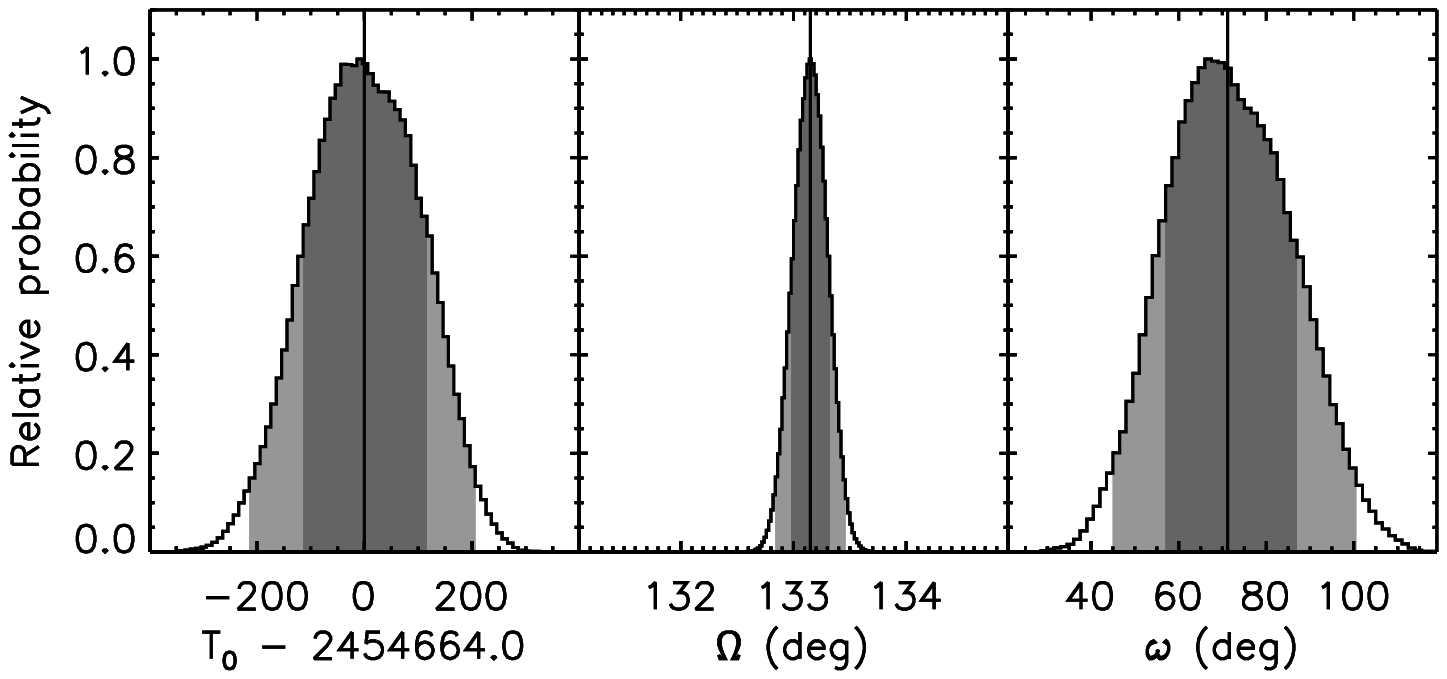}}

\caption{\normalsize Probability distributions of all orbital
  parameters derived from the MCMC analysis: semimajor axis ($a$),
  orbital period ($P$), eccentricity ($e$), inclination ($i$), epoch
  of periastron ($T_0$), PA of the ascending node ($\Omega$), and
  argument of periastron ($\omega$).  Each histogram is shaded to
  indicate the 68.3\% and 95.5\% confidence regions, which correspond
  to 1$\sigma$ and 2$\sigma$ for a normal distribution, and the solid
  vertical lines represent the median values.  Note that $T_0$ is shown
  in days since 2008 Jul 16 12:00 UT for
  clarity. \label{fig:orbit-parms}}

\end{figure}

Figure~\ref{fig:orbit-parms} shows the resulting MCMC probability
distributions for the seven orbital parameters of \hdbin, most of
which are somewhat non-Gaussian.  The two distributions that are most
nearly Gaussian and particularly well-constrained are the inclination
and the PA of the ascending node ($\Omega$), the latter of which along
with $\omega$ actually have a 180\degs\ ambiguity without radial
velocity information.  The best-fit parameters and their confidence
limits are given in Table~\ref{tbl:orbit}, and the best-fit orbit is
shown in Figure~\ref{fig:orbit-skyplot}.  The reduced $\chi^2$ of the
orbital solution is 1.14.

Applying Kepler's Third Law to the period and semimajor axis
distributions gives the posterior probability distribution for the
total mass of \hdbin, with a median of 0.1085~\Msun, a standard
deviation of 0.0018~\Msun, and 68(95)\% confidence limits of
$^{+0.0019}_{-0.0017}$($^{+0.004}_{-0.003}$)~\Msun\
(Figure~\ref{fig:orbit-mass}).  The MCMC probability distribution of
the total mass does not include the uncertainty in the parallax
(0.44\%), which by propagation of errors contributes an additional
1.3\% uncertainty in mass.  Since the MCMC-derived mass distribution
is asymmetric, we account for this additional error by randomly
drawing a normally distributed parallax value for each step in the
chain, which we then used to compute the distance and total mass.  The
resulting mass distribution is indistinguishable from Gaussian
(Figure~\ref{fig:orbit-mass}).  Our final determination of the total
mass is 0.109$\pm$0.002($^{+0.005}_{-0.004}$)~\Msun\ at 68(95)\%
confidence.  Thus, the total mass of this system is determined to 2\%
precision.

\begin{figure}
\vskip -1in
\hskip -0.3in
\centerline{\includegraphics[height=7.5in,angle=90]{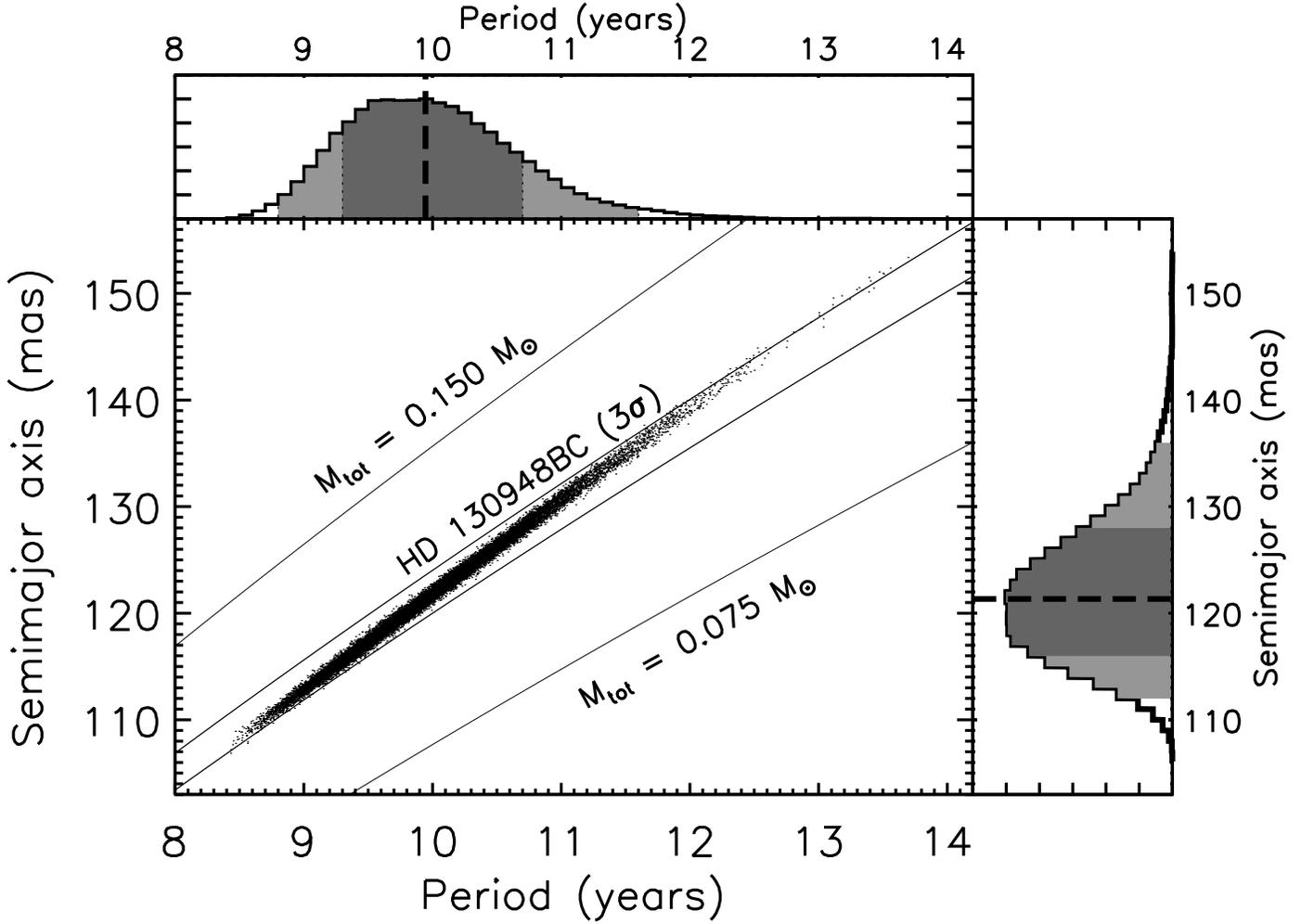}}
\vskip -2ex

\caption{\normalsize Results from the MCMC determination of the
  orbital period and semimajor axis for \hdbin.  The central plot
  shows all the values in the MCMC chain.  The locus illustrates the
  degeneracy between determining the orbital period and semimajor
  axis.  Lines of constant mass are drawn in the central plot to show
  that the resulting mass precision is much better than simply adding
  the uncertainties in $P$ and $a$ in quadrature.  The top and side
  plots show the resulting probability distributions of $P$ and $a$.
  Each histogram is shaded to indicate the 68.3\% and 95.5\%
  confidence limits, which correspond to 1$\sigma$ and 2$\sigma$ for a
  normal distribution, and the dashed vertical lines represent the
  median values. \label{fig:orbit-p_a}}

\end{figure}

\begin{figure}
\vskip -1in
\centerline{\includegraphics[height=7.5in,angle=90]{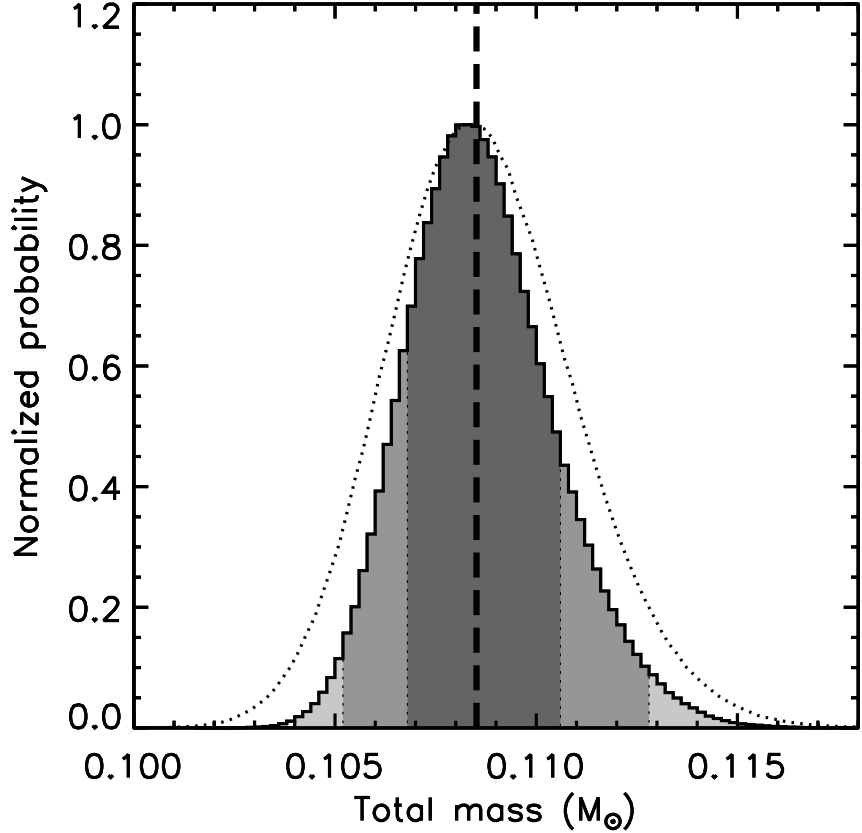}}
\vskip 2ex

\caption{\normalsize Probability distribution of the total mass of
  \hdbin\ resulting from our MCMC analysis.  The histogram is shaded
  to indicate the 68.3\%, 95.5\%, and 99.7\% confidence regions, which
  correspond to 1$\sigma$, 2$\sigma$, and 3$\sigma$ for a normal
  distribution. The dashed line represents the median value of
  0.1085~\Msun. The standard deviation of the distribution is
  0.0018~\Msun. The dotted unshaded curve shows the final mass
  distribution after accounting for the additional 1.3\% error due to
  the uncertainty in the \Hipparcos\ parallax of \hdprim; the result
  is essentially Gaussian.  The confidence limits for both
  distributions are given in
  Table~\ref{tbl:orbit}.  \label{fig:orbit-mass}}

\end{figure}

\begin{figure}
\vskip -1in
\centerline{\includegraphics[height=7.5in,angle=90]{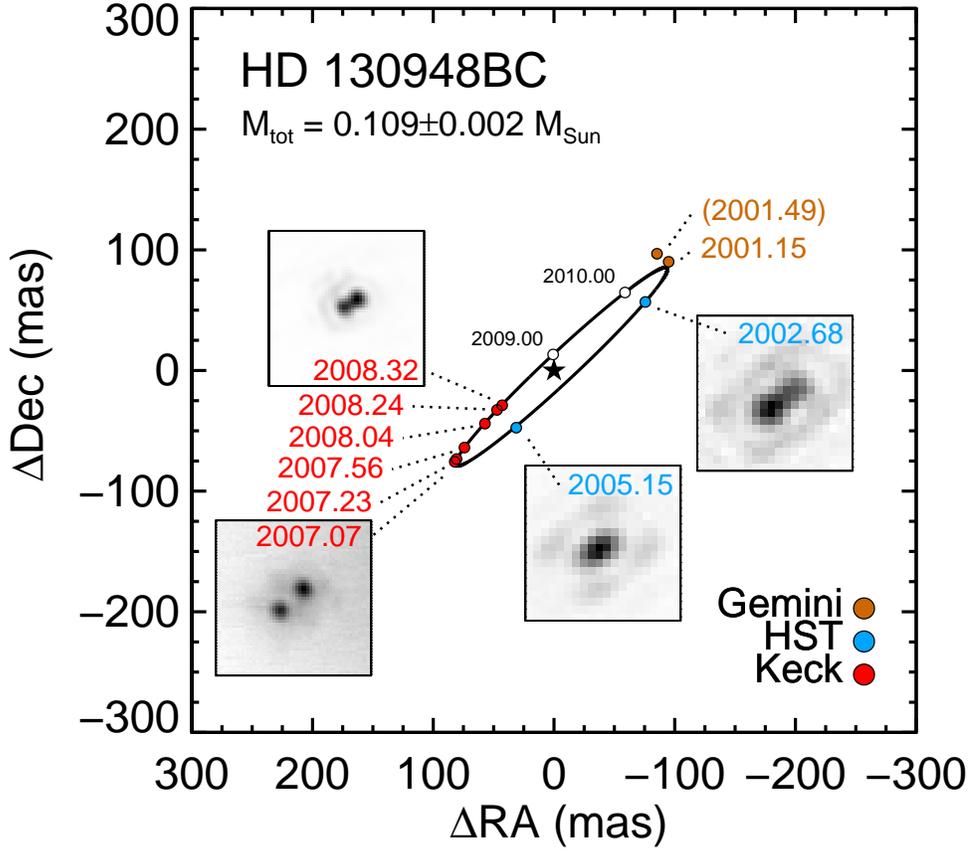}}

\caption{\normalsize Keck (red), \HST\ (blue), and Gemini (gold)
  relative astrometry for \hdbin\ along with the best-fit orbit using
  only the \HST\ and Keck data. The empty circles are the predicted
  location of HD~130948C for this object in 2009 and 2010. The Gemini
  point from 2001.15 was extracted from \citet{2003IAUS..211..265P}
  and seems to follow the best-fit orbit well. The Gemini point
  measured directly from archival data \citep[and consistent with the
  published values in][]{2002ApJ...567L.133P} is labeled in
  parentheses and is clearly discrepant from the best-fit
  orbit. \label{fig:orbit-skyplot}}

\end{figure}

The total mass is determined to a much higher precision than would be
calculated directly from the uncertainties in the orbital period and
semimajor axis because these two orbital parameters are strongly
correlated (Figure~\ref{fig:orbit-p_a}).  This is essentially a
consequence of encoding Kepler's Second Law (equal area per unit time)
in the orbit fitter, so that it naturally determines the ratio $a^2/P$
quite well.  Thus, the correlation between $a$ and $P$ roughly follows
lines of constant mass since $M_{tot}$ is just $(a^2/P)^2/a$.  This
property of orbit determination is well-known and has often been
utilized to measure dynamical masses even when the orbital parameters
are not well-constrained \citep[e.g.,][]{2003AJ....126.1971S}.

\subsection{Including Gemini Astrometry}

\begin{figure}
\vskip -1in
\hskip -0.75in \includegraphics[height=4.5in,angle=90]{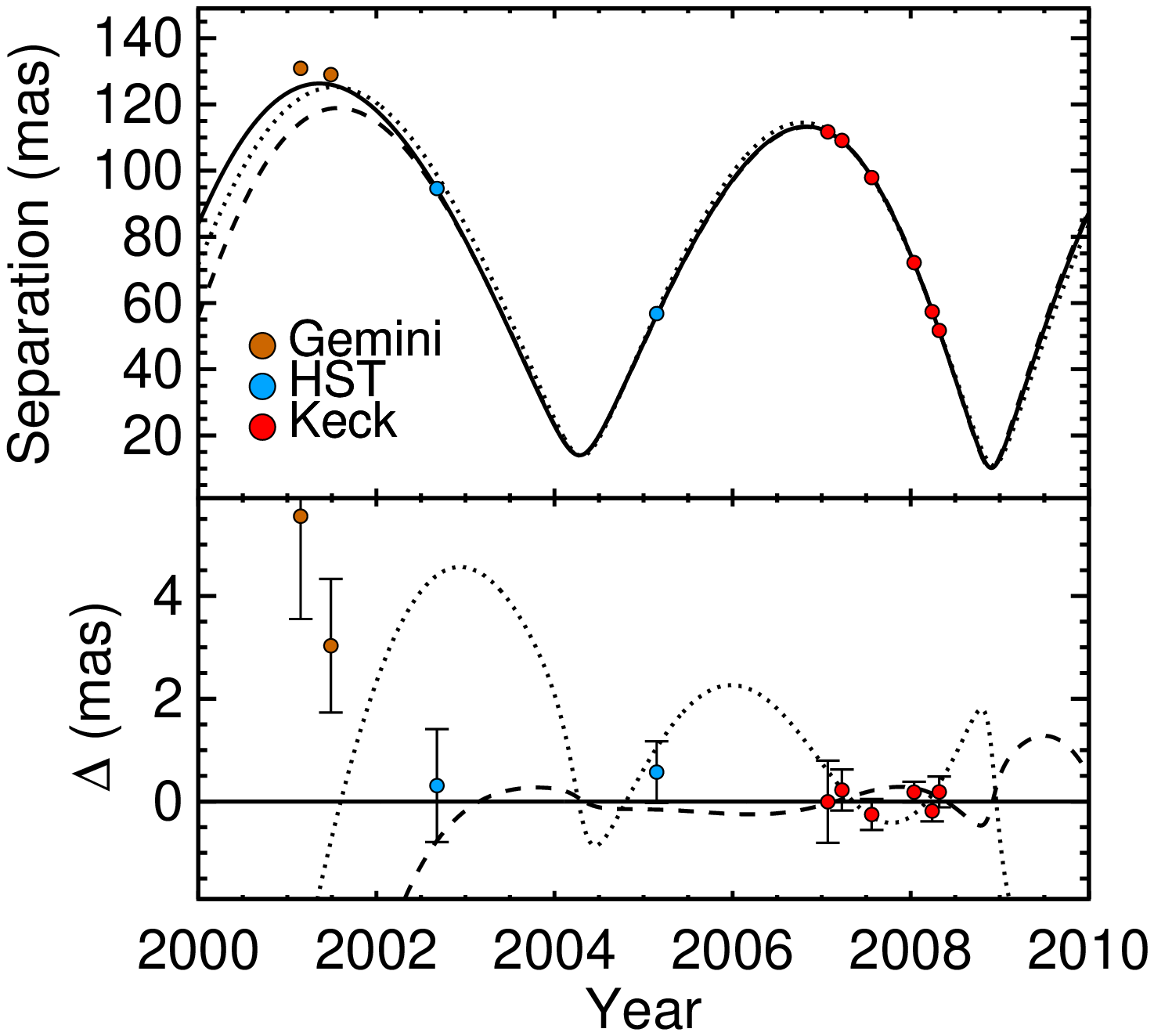}
\hskip -0.75in \includegraphics[height=4.5in,angle=90]{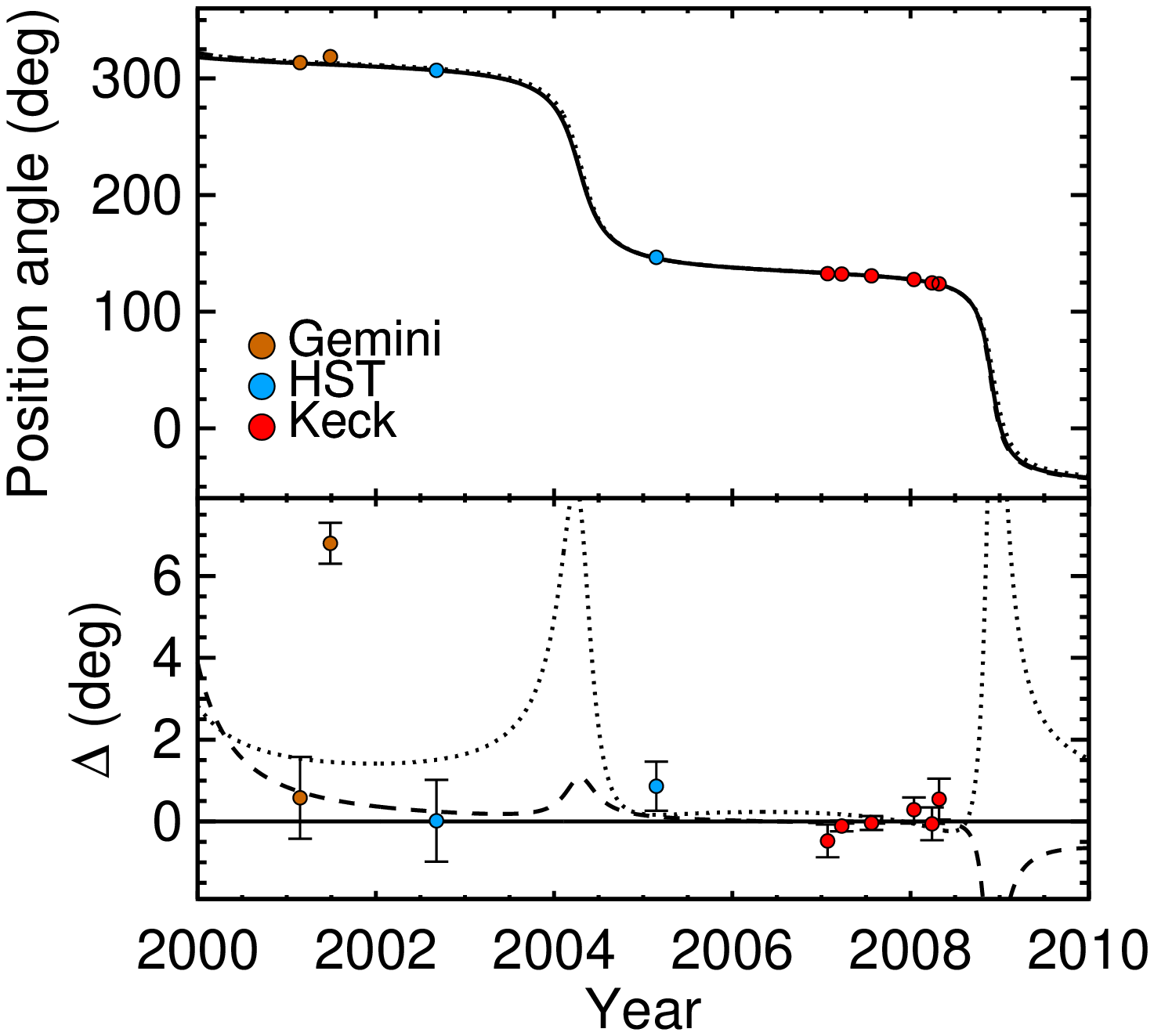}

\caption{\normalsize Keck (red), \HST\ (blue), and Gemini (gold)
  measurements of the separation ({\em left}) and PA ({\em right}) of
  \hdbin. The best-fit orbit is shown: as a solid line for \HST\ and
  Keck data only (our default solution); as a dashed line for \HST\
  and Keck data with the ``extracted'' Gemini data (2001.15 epoch);
  and as a dotted line for \HST\ and Keck data with the ``measured''
  Gemini data (2001.49 epoch). The bottom panels show the observed
  minus predicted separation and PA with observational error
  bars. This highlights the extreme discrepancy in the
  measured/published Gemini point in PA, even for the orbit in which
  it was included as a constraint (dotted). This shows that no
  physically plausible orbit can fit both the \HST+Keck data and the
  measured/published Gemini point, indicating a systematic error in
  the PA of the Gemini point.  There is a smaller discrepancy between
  the best-fit orbit and our ``extracted'' Gemini separation, possibly
  due to a systematic error in the instrument platescale in
  Wollaston-prism mode; however, the significance of the discrepancy
  is difficult to quantify because we cannot accurately assess the
  astrometric errors of the extracted data (we estimate 2~mas for the
  error in separation). \label{fig:orbit-sep-pa}}

\end{figure}

\begin{figure}
\vskip -1in
\centerline{\includegraphics[height=7.5in,angle=90]{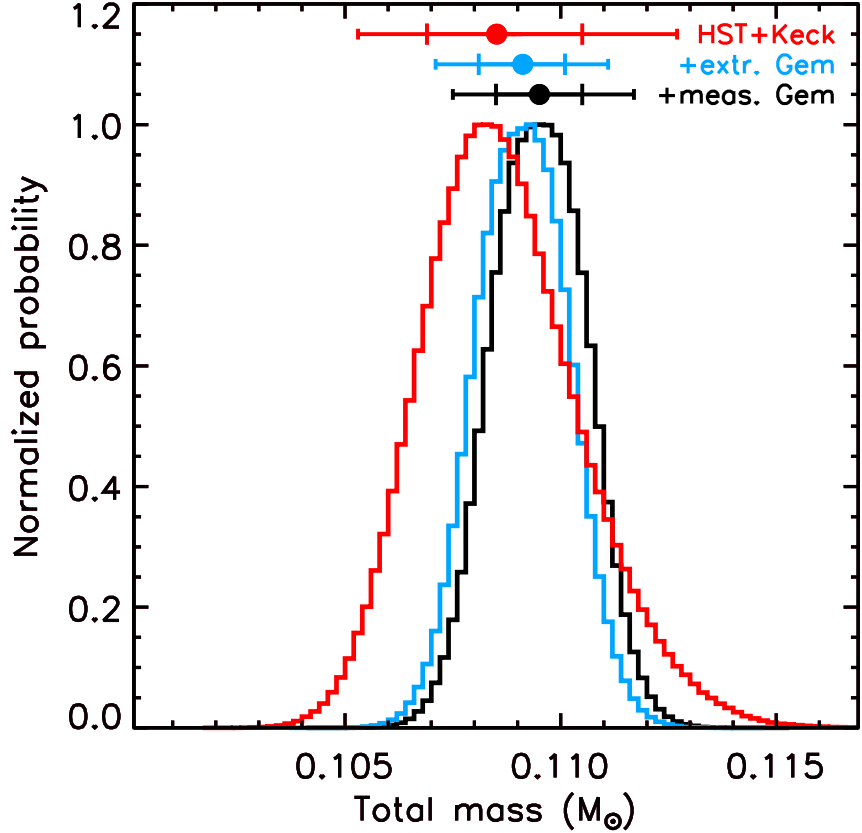}}
\vskip -4ex

\caption{\normalsize Total mass distribution from the MCMC analysis by
  fitting three different sets of astrometry: \HST\ and Keck data only
  (our default solution, {\em red}); \HST\ and Keck data with the
  ``extracted'' Gemini data ({\em gray}); \HST\ and Keck data with the
  ``measured'' Gemini data ({\em black}).  The filled circles indicate
  the median of the distributions, and the large (small) error bars
  indicate the 68.3\% (95.5\%) confidence limits, which correspond to
  1$\sigma$ (2$\sigma$) for a normal distribution.  Adding the
  ``extracted'' Gemini astrometry to the \HST\ and Keck data yields
  essentially the same dynamical mass but with a higher precision
  since it extends the time baseline of the observations.  Adding the
  ``measured'' Gemini astrometry also improves the nominal precision,
  but introduces a more significant systematic offset.  See
  \S~\ref{sec:gemini} for a discussion of the inconsistencies between
  the two Gemini measurements that cause this
  offset. \label{fig:compare-masses}}

\end{figure}

Our default best-fit orbit presented above includes only the \HST\ and
Keck measurements of \hdbin.  In section \S \ref{sec:gemini}, we
discussed the different Gemini measurements of \hdbin\ and their
inconsistencies.  The Gemini measurements have the potential to
improve the orbit determination by extending the observational time
baseline, so we explored the effect of adding each of these two
different measurements by running two additional MCMC chains.  One
chain uses the Gemini astrometry we extracted from
\citet{2003IAUS..211..265P} for the 2001~February~24~UT epoch, and the
other chain uses the Gemini astrometry we measured directly from
archival images at the 2001~June~28~UT epoch \citep[this astrometry is
consistent with that presented by][]{2002ApJ...567L.133P}.  For our
``extracted'' measurement, we estimated uncertainties of 2~mas in
separation and 1\degs\ in PA.  The MCMC chain with the addition of our
``extracted'' meaurement yielded a similar reduced $\chi^2$ (1.05) to
our default chain, and the orbital parameters were generally better
constrained (e.g., Figure~\ref{fig:compare-masses} shows that the
posterior mass distribution was somewhat tighter).  But the MCMC chain
with the addition of our ``measured'' Gemini astrometry had an
unacceptably large reduced $\chi^2$ (14.7), and the resulting orbital
parameters were generally inconsistent with our default chain.  The PA
of the ``measured'' Gemini point is grossly inconsistent with any
best-fit orbit: for instance, the binary must revolve backwards (with
respect to the Keck and \HST\ data) to be consistent with the
``measured'' Gemini point (Figure~\ref{fig:orbit-sep-pa}).  In other
words, the ``measured'' Gemini astrometry is inconsistent with any
physically plausible orbit that is also consistent with the \HST\ and
Keck astrometry.  Thus the inconsistencies between the ``extracted''
and ``measured'' astrometry discussed in \S \ref{sec:gemini} seem to
be due to a large systematic error in the PA of the ``measured''
Gemini astrometry (e.g., due to an incorrectly recorded orientation of
the detector).  While our ``extracted'' astrometry seems to improve
the orbit determination, we conservatively exclude it from our default
orbital solution as we have no way to accurately quantify its
astrometric uncertainties.


\section{Age of the Primary \hdprim \label{sec:age}}

Age determinations for individual main-sequence field stars are
challenging and imperfect.  Estimates generally rely on the slowing of
the stellar rotation period as stars grow older
\citep{1972ApJ...171..565S}.  Stars spin down as they age because
stellar winds carry away angular momentum; the slower rotation periods
then lead to a decline in stellar activity due to the underlying
stellar dynamo.

\citet{1998PASP..110.1259G} assigned an age range of 0.2--0.8~Gyr to
characterize his young solar-analog sample, of which \hdprim\ is a
member.  Given the importance of the system's age in interpreting our
mass measurement of the brown dwarf binary, we examine here the
specific properties of \hdprim\ to refine the age estimate.


\subsection{Chromospheric Activity \label{sec:age-rhk}}

For solar-type stars, chromospheric activity as traced by
\ion{Ca}{2}~HK emission provides one method to estimate ages.
\citet{don93, 1998csss...10.1235D} provide a calibration for this
index:
\begin{equation}
\log(t) = 10.725 - 1.334R_5 + 0.4085R_5^2  - 0.0522R_5^3
\end{equation}
where $t$ is the age in years and $R_5 = 10^5 R\arcmin_{HK}$, valid
for $\log(R\arcmin_{HK}) $ = $-$4.25 to $-$5.2.
\citet{1996AJ....111..439H} and \citet{2004ApJS..152..261W} measure
$\log(R\arcmin_{HK})$~=~$-$4.45 and $-$4.50 for \hdprim, respectively,
which translate into ages of 0.6 and 0.9~Gyr.  A clear error estimate
is not available for this relation.

\citet{mam08-ages} have updated this relation, incorporating new
samples of \ion{Ca}{2}~HK data, revised ages and membership lists for
nearby open clusters, and corrections for trends in activity with
stellar color (mass).  Their relation differs most notably from the
Donahue one at the youngest ages ($\lesssim$0.1~Gyr).  They find:
\begin{equation}
  \log(t) = -38.053 - 17.912\log(R\arcmin_{HK}) - 1.6675\log(R\arcmin_{HK})^2.
\end{equation}
The resulting implied ages for \hdprim\ are 0.4 and 0.6~Gyr for the
aforementioned $\log(R\arcmin_{HK})$ values.  \citet{mam08-ages}
estimate errors of $\approx$0.25~dex in the age ($\approx$60\%), based
on the dispersion produced by their relation when applied to binary
stars and star clusters.  We adopt an age from this method of
0.5$\pm$0.3~Gyr.

Direct comparison to the open cluster data used by \citet{mam08-ages}
provides useful reference points.  The Hyades
\citep[625~Myr;][]{1998A&A...331...81P} provides the most populous
sample for comparison; the cluster's median
$\log(R\arcmin_{HK})$~=~$-$4.47$\pm$0.09 (68\% confidence range) is
very well-matched to \hdprim.  The slightly younger UMa
\citep[500~Myr;][]{2003AJ....125.1980K} and Coma Ber
\citep[600~Myr;][]{2005PASP..117..911K} clusters also have comparable
values of $-$4.48$\pm$0.09 and $-$4.43 (no confidence limits given),
respectively, though with much smaller samples of stars.\footnote{Note
  that the ages for these latter two clusters are older than adopted
  in the Donahue analysis.  Combined with the larger sample of young
  clusters, the behavior of the activity--age relation from
  \citet{mam08-ages} produces the somewhat younger age for \hdprim\
  compared to \citep{1998csss...10.1235D}.}  The data for \hdprim\ are
clearly inconsistent with older clusters NGC~752
\citep[2~Gyr;][]{1995AJ....109.2090D} and M~67
\citep[4~Gyr;][]{1999AJ....118.2894S, 2004PASP..116..997V} that have
$\log(R\arcmin_{HK})$ values of $-$4.70 and $-$4.84$\pm$0.11,
respectively.  However, the activity data for \hdprim\ are still
formally consistent with ages as young as the Pleiades
($-$4.33$\pm$0.24), given the large scatter in the sample for that
cluster.


\subsection{X-Ray Emission \label{sec:age-xray}}

X-ray emission of solar-type stars also declines with age.
\citet{1999A&AS..135..319H} measure $\log(L_X)$~=~29.0~dex (cgs) for
\hdprim\ with a 7\% uncertainty.  \citet{1998PASP..110.1259G} provides
an age calibration based on scaling relations for stellar activity:
\begin{equation}
\log(L_X/L_{\rm bol}) = -6.38 - 2.6\alpha\log(t_9/4.6) + \log[1 + 0.4(1-t_9/4.6)]
\end{equation}
where $t_9$ is the age in Gyr and $\alpha$ is the coefficient that
relates rotation period to stellar age, either $\alpha$~=~0.5
\citep{1972ApJ...171..565S} or $\alpha~=~1/\exp$ \citep{wal91}.
Following \citet{wils01}, we adopt the zero-point of $-$6.38 based on
the X-ray luminosity of the Sun from \citet{1987ApJ...315..687M}.
Adopting the \Lbol/\Lsun = 1.21 from \citet{1998PASP..110.1259G} gives
$\log(L_X/L_{\rm bol})$~=~$-$4.70 for \hdprim, which corresponds to an
estimated age of 0.1--0.3~Gyr, depending on the value of $\alpha$.

\citet{mam08-ages} find that X-ray emission is strongly correlated
with $\log(R\arcmin_{HK})$, and they derive a relation between X-ray
activity and age from this correlation and their relation between
chromospheric activity and age:
\begin{equation}
\log(t) = 1.20 - 2.307\log(L_X/L_{\rm bol}) - 0.1512\log(L_X/L_{\rm bol})^2
\end{equation}
where $t$ is the age in years.  This relation gives an age of 0.5~Gyr
for \hdprim, which perhaps not surprisingly is in agreement with the
age estimate from their chromospheric activity relation
(Equation~2).  This agreement indicates that \hdprim\ shows typical
X-ray emission given its chromospheric activity level.

As a more direct point of reference, the X-ray luminosity of \hdprim\
in excellent agreement with single G~stars in the Pleiades and Hyades,
where the average values are 28.9--29.0 \citep{1995ApJ...448..683S,
  2001A&A...377..538S}.  Unfortunately, there is only a modest
difference in the distribution of X-ray luminosities for G~stars in
these two clusters \citep[e.g.,][]{2005ApJS..160..390P}.  Similarly,
\citet{2003ApJ...585..878K} have noted that the character of stellar
X-ray emission changes with age, with older stars having softer
\ROSAT\ X-ray emission.  The \ROSAT\ X-ray hardness ratios $HR1$ and
$HR2$ for \hdprim\ are $-$0.34$\pm$0.07 and $-$0.08$\pm$0.012,
respectively \citep{1999A&A...349..389V}, values which are in good
agreement with G~stars in the Hyades and largely distinct from young
moving group members ($\approx$10--30~Myr) and nearby (old) stars.
Overall, the X-ray data for \hdprim\ support an age around the
Pleiades and Hyades clusters (i.e., 125--625~Myr) but do not provide a
more definitive age estimate.


\subsection{ Rotation/Gyrochronology \label{sec:age-gyro}}

\citet{2007ApJ...669.1167B} proposed a method for determining the ages
of solar-type main-sequence stars based on stellar rotation
(``gyrochronology'').  This technique is potentially the most direct
and precise method for estimating stellar ages.  Over a star's
lifetime, stellar winds carry away angular momentum, so the star
rotates more slowly as it ages.  The functional form of the spin-down
was found by \citet{1972ApJ...171..565S} to be proportional to
$\sqrt{t}$.  \citet{2007ApJ...669.1167B} adds a separable color
dependence, which translates to a mass dependence, to the spin-down.
This is empirically motivated and suggests a mass dependence on the
angular momentum loss, and thus on the strength of the stellar dynamo.
In fact, his empirical relation cannot be used for all stars, but only
those which share a common dynamo mechanism, one that is presumed to
originate at the interface between convective and radiative zones in
the stellar interior.  Fully convective stars are expected to have a
different dynamo mechanism that is weaker and prevents efficient
spin-down, leading these stars to rotate rapidly
($P$~$\lesssim$~2~days).  This interpretation is empirically motivated
by stellar rotation data which show two sequences of stars, with some
stars in transition between the two states.  If a main-sequence star
can be shown to be part of the ``interface sequence'' (as almost all
stars older than about 200~Myr seem to be), then its rotation period
and $B-V$ color can be used to determine its age to a precision of
15--20\% using the empirical gyro relation of
\citet{2007ApJ...669.1167B}.

\citet{2000AJ....120.1006G} measured rotational modulation of \hdprim\
from photoelectric time series photometry spanning 188~days.  He found
two distinct but similar periods of 7.69~days and 7.99~days, so we
adopt the mean and standard deviation as the rotation period for
\hdprim\ and its uncertainty ($P$~=~7.84$\pm$0.21~days).  For the
$B-V$ color, we adopt 0.576$\pm$0.016~mag from the \Hipparcos\ catalog
\citep{1997A&A...323L..49P}.

First, we show that we are justified in applying the
\citet{2007ApJ...669.1167B} relation to \hdprim\ because it seems to
have an interface dynamo.  We have verified this in two independent
ways: (1) its age-normalized rotation period ($P/\sqrt{t}$, where $t$
is 0.2--0.8~Gyr, the age estimated from stellar activity indicators)
and $B-V$ color place it within the ``interface sequence'' of stars in
open clusters \citep[see Figure 1 of][]{2007ApJ...669.1167B}; and (2)
its X-ray flux ($\log{L_X/L_{\rm bol}}$~=~-4.70 from \S
\ref{sec:age-xray}) and Rossby number (defined as $P/\tau_c$~=~0.29,
where $\tau_c$ is the convective turnover timescale of \hdprim,
estimated to be 27~days according to \citealp{1996ApJ...457..340K})
clearly place it among other X-ray active stars with interface dynamos
\citep[see Figure 1 of][]{2003ApJ...586L.145B}.

The functional form of the \citet{2007ApJ...669.1167B}
gyrochronological age relation is:
\begin{equation}
\log(t) = \frac{1}{n}[\log(P) - \log(a) - b\log(B-V-c)]
\end{equation}
where $t$ is the age in years, $P$ is the rotation period in days, $n$
is the power-law exponent of the spin-down, and $a$, $b$, and $c$ are
empirical coefficients.  By fitting rotation periods and colors for
stars in open clusters of known ages, \citet{2007ApJ...669.1167B}
found coefficients of $a$~=~0.7725$\pm$0.011, $b$~=~0.601$\pm$0.024,
$c$~=~0.4 (not free to vary), and $n$~=~0.5189$\pm$0.0070. We have
employed the gyro relation in a Monte Carlo fashion, drawing the
observational inputs and functional coefficients from normal
distributions consistent with their quoted errors.  This approach
yields a standard deviation in the resulting age distribution
consistent with the error one would obtain by propagation of errors
\citep[i.e., Equation 11 of][]{2007ApJ...669.1167B}, but it preserves
asymmetries in the resulting confidence limits.  We thereby find a
gyro age of 0.65$^{+0.13(0.28)}_{-0.10(0.18)}$~Gyr for \hdprim\
(68(95)\% confidence limits).

\citet{mam08-ages} have derived new coefficients for the gyro relation
that improve its agreement with observations of the Hyades, Pleiades,
and the Sun: $a$~=~0.407$\pm$0.021, $b$~=~0.325$\pm$0.024,
$c$~=~0.495$\pm$0.010, and $n$~=~0.566$\pm$0.008.  Using these, we
find a gyro age of 0.79$^{+0.22(0.53)}_{-0.15(0.26)}$~Gyr for \hdprim\
(68(95)\% confidence limits).  This is the age we adopt as the
gyrochronological age of \hdprim, despite its somewhat larger
uncertainties, which stem from the larger uncertainties in the
coefficients determined by \citet{mam08-ages}.  The age estimates
derived using the different sets of coefficients are consistent with
each other, though the improved coefficients give a slightly older
age.  Because of the scarcity of rotation period data for clusters, no
stars older than the Hyades except the Sun were used in either
calibration of the gyrochronology relation.

Comparison of \hdprim's properties to the stellar rotation data
presented by \citet{mam08-ages} offers a direct assessment of its
gyrochronolgical age.  For an object of its color, it appears to be in
good agreement with members of the Hyades
\citep[625~Myr;][]{1998A&A...331...81P}.  In fact, its rotation period
is slightly slower than the mean Hyades rotation period. This implies
an age slightly older than, but marginally consistent with, the
Hyades, which supports our adopting the gyro age of \hdage.


\subsection{Other Age Indicators \label{sec:age-other}}

The location on the H-R diagram relative to stellar evolutionary
isochrones provides another age estimate, though for main-sequence
stars this is limited since stellar luminosity changes gradually with
age.  Using high resolution spectroscopic data combined with
bolometric magnitudes and isochrones, \citet{2005ApJS..159..141V}
derive an age estimate of 1.8~Gyr with a possible range of
0.4---3.2~Gyr.  From the same data and with more detailed analysis,
\citet{2006astro.ph..7235T} infer a median age of 0.72~Gyr with a 95\%
confidence range of 0.32--2.48~Gyr.

For solar-type stars, photospheric lithium is depleted with age, as
indicated from \ion{Li}{1}~$\lambda$6708 measurements for stars in
open clusters, with modest changes for $\lesssim$100~Myr and then more
rapid depletion with age.  However, even within a given cluster,
lithium abundances show a substantial scatter for stars of a given
color (mass), and thus we consider the Li data only as a qualitative
check.  Measurements by \citet{1981ApJ...248..651D},
\citet{1985ApJ...290..284H}, and \citet{2001A&A...371..943C} give
\ion{Li}{1}~$\lambda$6708 equivalent widths of 95$\pm$14, 96$\pm$3,
and 103.1$\pm$3~m\AA, respectively, for \hdprim.  Compared to stars of
similar $B-V$~=~0.58, these values are slightly lower than the mean
for the Pleiades and slightly higher than for UMa and the Hyades,
though consistent with the scatter in each cluster's measurements
\citep{1993AJ....106.1080S, 1993AJ....106.1059S, 1993AJ....105.2299S}.

\citet{2000AJ....120.1006G} examined the space motions of his young
solar analog sample and did not associate \hdprim\ with any of the
known moving groups.  Using his space motions, we confirm that
\hdprim\ does not belong to any moving groups that were identified
after his analysis \citep{2004ARA&A..42..685Z, 2006ApJ...649L.115Z,
  2006ApJ...643.1160L}.  Thus, the space motion of \hdprim\ offers no
constraint on its age.

Overall, the isochrone analysis and lithium abundances are consistent
with the activity-derived ages, albeit with lower precision.


\subsection{Age Summary \label{sec:age-summ}}

The age estimates for \hdprim\ are summarized in
Table~\ref{tbl:age}. The most precise estimates are derived from the
connection between stellar rotation (thus activity) and age.  Using
gyrochronology, we have estimated the age of \hdprim\ from its
rotation period to be \hdage.  This is consistent with the
0.5$\pm$0.3~Gyr age derived from the most up-to-date relation between
chromospheric activity and age \citep{mam08-ages}.  The larger
uncertainty in the activity age ($\approx$60\%) compared to the gyro
age ($\approx$25\%) makes it somewhat less attractive, although it is
better calibrated at ages intermediate between the Hyades the Sun.
Less precise estimates are available from X-ray activity, lithium
depletion, and stellar evolutionary isochrones, and these are all
consistent with the more precise gyro age estimate.

It is important to note that the rotation data for \hdprim\ are
generally inconsistent with ages younger than the Hyades, so we find
it unlikely that \hdprim\ is much younger than $\approx$0.6~Gyr
\citep[see Figure~10 of][]{mam08-ages}.  This assertion is free from
the uncertainties in the calibration of the gyro relation at older
ages, thus our 1$\sigma$ (2$\sigma$) lower limits on the gyro age of
0.64~Gyr (0.53~Gyr) are expected to be reasonable.  Additionally, the
chromospheric activity of \hdprim\ is inconsistent with stars in
clusters much older than the Hyades \citep[see Figure~4
of][]{mam08-ages}.  Thus, the ensemble of data can be made fully
consistent for an age of \hdprim\ that is roughly the same as the
Hyades.

In the analysis that follows, we adopt the \hdage\ age estimate from
gyrochronology for \hdprim, not only because it provides the best
precision, but also because it is the most fundamental age indicator
available.  It directly probes stellar angular momentum loss, whereas
activity indicators indirectly probe the change in stellar rotation
through its impact on the stellar dynamo.  However, we are reluctant
we caution that the gyro age still has at least two potential
uncertainties: (1) the rotation period was measured over only one
season and so is not completely irreproachable;\footnote{The rotation
  period measured from photometric modulation due to star spots can be
  affected by the latitude of the spots since there is likely to be
  some amount of differential rotation.  The fact that
  \citet{2000AJ....120.1006G} measured two similar but distinct
  rotation periods for \hdprim\ highlights the difficulties in
  deriving a robust rotation period from time series photometry.} (2)
the gyrochronological age relation is not well-calibrated at ages
older than the Hyades, as the only age datum $\gtrsim$0.6~Gyr is the
Sun.  For these reasons, we will also consider the activity age of
0.5$\pm$0.3~Gyr, which is less precise but is somewhat better
calibrated at ages $\gtrsim$0.6~Gyr.  Future asteroseismology
measurements that probe the interior structure of \hdprim\ could
provide an even more fundamental and precise age estimate.  However,
such measurements have only been obtained for a handful of very bright
stars to date \citep{2002A&A...394L...5F, 2004A&A...417..235E,
  2004A&A...418..295M, 2005A&A...434.1085C, 2005A&A...440..609B,
  2006A&A...449..293E}.


\section{Discussion}

The direct measurement of the masses and/or ages of ultracool dwarfs
is one of the few avenues by which theoretical models describing these
objects can be tested.  \hdbin\ is unique among ultracool dwarf
binaries with dynamical mass determinations to date because the
primary \hdprim\ offers an independent age and metallicity constraint,
under the conservative assumption that all three components are coeval
and have the same composition.  The metallicity of \hdprim\ is
[M/H]~=~0.00 \citep{2005ApJS..159..141V}, so the publicly available
solar-metallicity models are well-suited to our analysis.  Because the
age of \hdprim\ is constrained to a lower precision ($\approx$25\%)
than the total mass of \hdbin\ (2\%), we have conducted the model
comparisons discussed here with minimal dependence on the measured age
of \hdprim\ so that future improvements in the age measurement can be
readily applied to our results.

In the following analysis, we utilized all available measurements of
\hdbin: the total mass from this work; the Keck $JHK$ photometry; the
\Hipparcos-measured distance; and the individual spectral types.  We
randomly drew the measured properties of \hdbin\ from appropriate
distributions, carefully accounting for the covariance between the
different quantities.  For example, $M_{tot}$ and \Lbol\ are
correlated through the distance (this has a small effect on our
analysis), and the luminosities of the two components are correlated
because the flux ratio is measured to higher accuracy than the total
flux.

We have deliberately chosen to use \Lbol\ rather than \Teff\ as the
basis of our model comparisons because values of \Teff\ in the
literature are invariably tied to either evolutionary or atmospheric
theoretical models in some way.  By using \Lbol, which only depends on
direct measurements of SEDs and distances, we have avoided such
circular comparisons.  In the following, we consider two independent
sets of evolutionary models: the Tucson models
\citep{1997ApJ...491..856B} and the Lyon DUSTY models
\citep{2000ApJ...542..464C}, which are appropriate for mid-L dwarfs
such as \hdbin\ with significant amounts of dust in their
photospheres.

\subsection{Spectral Types \label{sec:spt}}

\citet{2002ApJ...567L.133P} originally used resolved $J$-band spectra
to find spectral types of dL2$\pm$2, which are on the
\citet{1999AJ....118.2466M} system, for both components of \hdbin.  In
addition to large quoted uncertainties, this measurement suffers from
systematic errors inherent to AO-fed slit spectroscopy.  The quality
of the AO correction is wavelength dependent, and this ultimately
leads to a modification of the shape of the continuum.  Since spectral
typing of brown dwarfs is largely based on matching the continuum to
spectral standards, this modification must be corrected first, which
\citet{2002ApJ...567L.133P} did not do.

\citet{2002ApJ...567L..59G} determined that the spectral types of
HD~130948B and C are indistinguishable by matching the $H$- and
$K$-band resolved spectra with spectral templates.  Before matching
the spectra, they first applied an empirical correction to the
continuum shape.  The near-infrared spectral type of their best
matching template spectrum (2MASS~J0036+1821) is L4$\pm$1
\citep{2004AJ....127.3553K}, and its optical spectral type is L3.5
\citep{2000AJ....120..447K}.  Since the template matching was done in
the near-infrared, we follow \citet{2002ApJ...567L..59G} in adopting
spectral types of L4 for both components of \hdbin.

In principle, the spectral template matching technique employed by
\citet{2002ApJ...567L..59G} had relative precision of 0.5 subclasses,
since they compared template spectra at each integer subclass.
However, we adopt an uncertainty in the spectral type of $\pm$1 since
their method relies on a non-standard technique, and the spectra were
matched in the near-infrared where the best-fitting template
(2MASS~J0036+1821) has a spectral type uncertainty of $\pm$1. In the
following analysis, we treat the errors in the spectral types of the
two components as independent, such that $\Delta$SpT~=~0.0$\pm$1.4.

\subsection{Bolometric Luminosities \label{sec:lbol}}

We calculated the individual bolometric luminosities of \hdbin\ by
using $K$-band bolometric corrections from the BC$_K$--SpT relation of
\citet{gol04} and our $K$-band absolute magnitudes.  We added in
quadrature the error in $M_K$ (0.03~mag), the error resulting from the
$\pm$1 subtype uncertainty in spectral classification (0.015~mag), and
the RMS scatter in the BC$_K$--SpT polynomial relation (0.13~mag).
Thus, our derived values of \Lbol\ for \hdbin\ have uncertainties of
0.05~dex (Table~\ref{tbl:meas}).  These uncertainties may be improved
by future direct measurements of the resolved SEDs of \hdbin, since
the RMS scatter in the BC$_K$--SpT relation dominates the errors.

The ratio of the bolometric luminosities of \hdbin\ is known more
precisely than the individual values.  This is because the error in
the luminosity ratio does not include the error in distance and
intrinsic scatter in the BC$_K$--SpT relation that are common to both
components.  We therefore combined the weighted average of our
measured $K$-band flux ratios ($\Delta{K}$~=~0.197$\pm$0.008~mag) with
the expected difference in bolometric correction ($\Delta{{\rm
    BC}_K}$~=~0.00$\pm$0.02~mag; where the uncertainty is due to the
independent error on the spectral type of each component) to derive a
luminosity ratio of $\Delta\log$(\Lbol)~=~0.079$\pm$0.008~dex.  In the
following analysis, we treat the individual luminosities of \hdbin\ as
correlated, in order to preserve the precision in the luminosity
ratio.  Thus, we are able to determine relative quantities (e.g., the
mass ratio and $\Delta\Teff$) to much higher precision than if we
incorrectly treated the 0.05~dex uncertainties in \Lbol\ as
independent.

\subsection{Model-Inferred Age \label{sec:modelage}}

Brown dwarf model cooling sequences are usually thought of as
predicting an observable quantity (\Teff, \Lbol, etc.) from the pair
of fundamental parameters mass and age.  Here we have measured the
total mass and individual luminosities, so we instead use these two
quantities to infer the third: age.  In other words, brown dwarf
cooling sequences define a mass--luminosity--age relation, so the
measurement of any two of these specifies the third.  As suggested by
\citet{2008arXiv0807.0238L}, ``mass benchmarks'' like \hdbin\ can
offer even tighter constraints on model-inferred properties than ``age
benchmarks'' that are more often considered in the literature.

\begin{figure} 
\includegraphics[height=6.5in,angle=90]{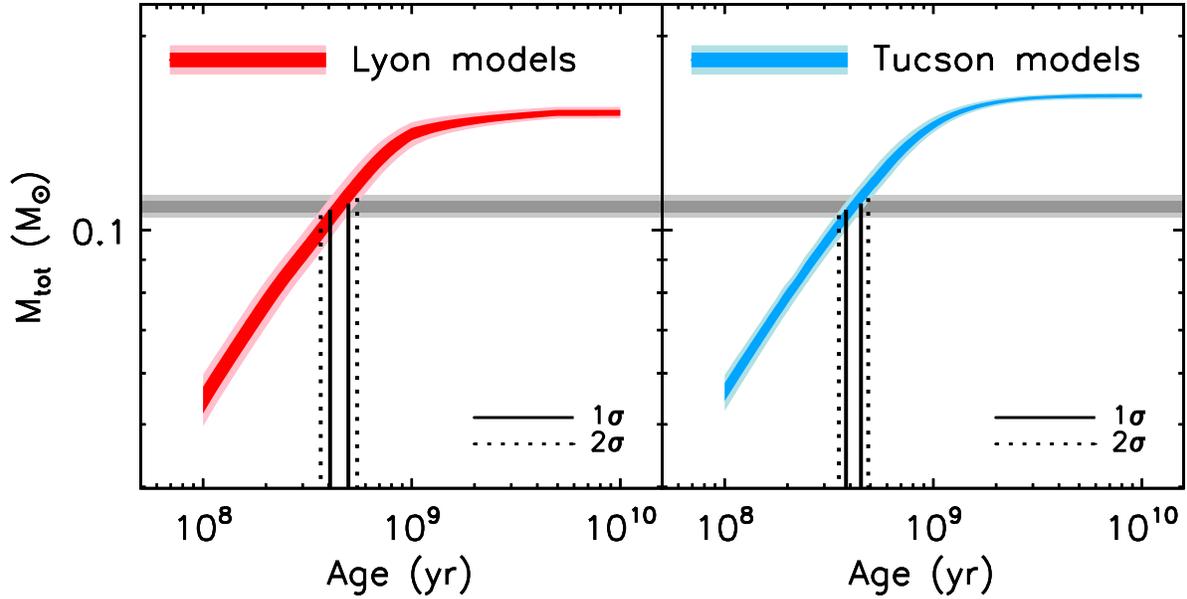}

\caption{ \normalsize Total mass of \hdbin\ as a function of the age
  of the system that is predicted by evolutionary models, given the
  observed luminosities of the two components.  By applying the
  measured total mass ($M_{tot}$), we inferred the age of \hdbin\ from
  evolutionary models (see \S \ref{sec:modelage}).  The colored shaded
  regions indicate the 1$\sigma$ and 2$\sigma$ ranges in $M_{tot}$
  corresponding to the luminosity uncertainties. At older ages, model
  substellar objects must be more massive in order to match the
  imposed luminosity constraint.  At the oldest ages, the measured
  luminosities of \hdbin\ would correspond to a star at the bottom of
  the main-sequence, which causes the flattening of the $M_{tot}$--age
  curves.  The horizontal gray bars show our 1$\sigma$ and 2$\sigma$
  constraints on the total mass. The intersection of the measured
  $M_{tot}$ with the model-predicted $M_{tot}$ is shown by the solid
  (dotted) lines and corresponds to our 1$\sigma$ (2$\sigma$) derived
  age range. Note that the model tracks shown here correspond only to
  objects with the same individual luminosities as HD~130948B and C
  and are not generally applicable to other
  binaries. \label{fig:mtot-age}}

\end{figure}

At each model age, the individual luminosities we have measured fully
determine the model-predicted individual masses.  Thus, we use the
evolutionary models to calculate model masses $M_{\rm B}$ and $M_{\rm
  C}$ as a function of age, and this yields the model-predicted
$M_{tot}$ as a function of age.  We then impose the constraint of the
observed $M_{tot}$, which uniquely determines the model age.  We
perform this calculation many times using randomly drawn total masses
and individual luminosities to simulate the observational
uncertainties while accounting for the covariance due to the distance
error.  The median and standard deviation of the resulting age
distribution is given in Table~\ref{tbl:model}. Given the very precise
mass measurement (2\%), the $\approx$13\% error on the luminosity
dominates the resulting uncertainty in the model-inferred age.  This
procedure for inferring the age of \hdbin\ from evolutionary models is
depicted in Figure~\ref{fig:mtot-age}.

The age inferred from Tucson models, 0.41$^{+0.04}_{-0.03}$~Gyr, is
slightly younger than inferred from the Lyon models,
0.45$^{+0.05}_{-0.04}$~Gyr, though both model-inferred ages agree
within the errors.  The combination of theoretical models with
measurements of the total mass and luminosities of \hdbin\ yields
extremely small uncertainties ($\lesssim$10\%) in the age of the
HD~130948 system.  The Lyon and Tucson model-inferred ages are
2.4$\sigma$ and 2.2$\sigma$ discrepant, respectively, with the \hdage\
age estimate for \hdprim.  Since the model-inferred age was derived
from the observed total mass and individual luminosities of \hdbin,
this disagreement is essentially a statement of how well (or not)
models predict the luminosity evolution of objects with the masses of
\hdbin.  We discuss this discrepancy further in \S
\ref{sec:lbol-evol} and \S \ref{sec:teff-hr}.

\subsubsection{Comparison to L~Dwarfs in Clusters and Moving Groups \label{sec:jameson}}

\citet{2008MNRAS.tmp..282J} have proposed a method for estimating the
ages of young ($\lesssim$0.7~Gyr) L~dwarfs for which the $J-K$ color
and $K$-band absolute magnitude ($M_K$) are known.  Their empirical
relation is calibrated by L~dwarfs in clusters and moving groups,
whose ages have been determined in the literature from model stellar
isochrones.  Applying their empirical relation and accounting for the
error in the absolute magnitudes and colors, we derive a median age
and 1$\sigma$ (2$\sigma$) confidence limits of
0.25$^{+0.06(0.12)}_{-0.06(0.11)}$~Gyr for both components of \hdbin.
It is worth noting that the empirical relation gives identical results
for the two components even though the measured colors and absolute
magnitudes of the two components are essentially independent.

The age derived from the \citet{2008MNRAS.tmp..282J} empirical
relation is systematically younger by about 0.2~Gyr than the
model-inferred age of \hdbin.  This 2--3$\sigma$ disagreement is
perhaps not surprising since for ages older than that of the Pleiades
($>$~125~Myr), the empirical relation is constrained by only 3
L~dwarfs in the Hyades (625~Myr) and 2 L~dwarfs in the Ursa Major
moving group, for which they adopt an age of 400~Myr.  As discussed by
\citet{2008MNRAS.tmp..282J}, the age estimate of the Ursa Major moving
group varies in the literature from 0.3--0.6~Gyr.\footnote{For
  example, \citet{mam08-ages} adopt an age of 500~Myr for this
  association.  Therefore, in \S \ref{sec:age-rhk} we have also
  implictly adopted this age.} We speculate that if an older age had
been adopted for the Ursa Major moving group, instead of 0.4~Gyr, the
empirical age relation would likely yield an older age more consistent
with that inferred from evolutionary models.  However, the estimated
age of \hdprim\ (\hdage) is even older than that inferred from
evolutionary models.  This creates a much larger discrepancy
($>$3$\sigma$) with the 0.25~Gyr age derived from the
\citet{2008MNRAS.tmp..282J} empirical relation.

If \hdprim, and thus \hdbin, is indeed this old, then the empirical
relation would not be applicable to this system.  Direct examination
of the color-magnitude diagram of their sample \citep[Figure~1
of][]{2008MNRAS.tmp..282J} shows that the region populated by Hyades
and Ursa Major L~dwarfs also contains many field L~dwarfs.  These
objects could indeed be young, or they could be older field dwarfs
whose color-magnitude evolution has brought them into that region of
the diagram.  In fact, {\em all} field L~dwarfs
($M_K$~$\gtrsim$~13~mag) seem to lie in a region of the
color-magnitude diagram that would imply ages intermediate between the
Pleiades and Hyades.  This fact and the inconsistency between our
derived age and that of \hdprim\ seem to highlight the danger of using
this empirical age relation for L~dwarfs older than its stated
applicable range ($<$0.7~Gyr).  Examination of L~dwarfs with
independent age estimates $>$0.7~Gyr would verify whether spuriously
young ages can be derived for L~dwarfs that are, in fact, older.

\subsection{Mass Ratio and Substellarity \label{sec:qratio}}

\begin{figure}
\centerline{\includegraphics[width=6.5in,angle=0]{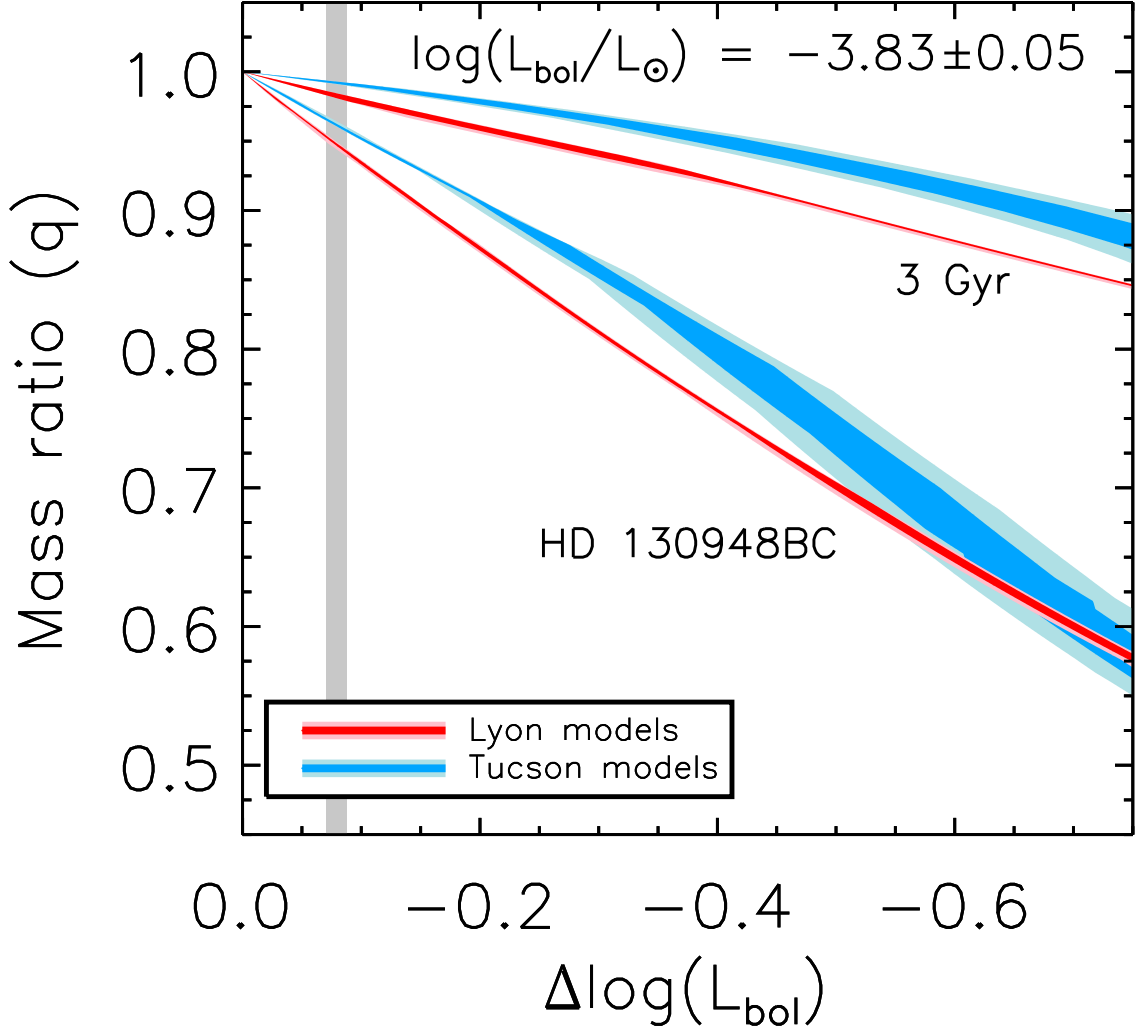}}

\caption{ \normalsize The model-predicted mass ratio for \hdbin\ as a
  function of the observed luminosity ratio.  The colored regions are
  the 1$\sigma$ and 2$\sigma$ ranges of possible mass ratios that
  correspond to the 1$\sigma$ and 2$\sigma$ uncertainties in the
  luminosity of HD~130948B at the model-inferred age.  The thin gray
  box shows the 1$\sigma$ range of the measured luminosity ratio of
  \hdbin.  A second set of colored regions shows the model-inferred
  mass ratios for an age of 3~Gyr, which illustrates the weak
  dependence of the assumed age on the mass ratio at near-unity flux
  ratios. Since the inferred mass ratio of a nearly equal-magnitude
  binary such as \hdbin\ is very insensitive both to the age of the
  system and to the evolutionary models used, the individual masses of
  \hdbin\ can be determined robustly. Note that these curves are not
  generally applicable to other binaries, since they are drawn for a
  small range in primary component \Lbol\ and system
  age.  \label{fig:mass-ratio}}

\end{figure}

By measuring the relative orbit of \hdbin, we have determined its
total mass very precisely. In order to calculate individual masses, we
must infer the mass ratio ($q$~$\equiv$~$M_{\rm C}/M_{\rm B}$) from
evolutionary models. Fortunately, the inferred mass ratio is very
weakly dependent on theoretical models given the near-unity flux ratio
of \hdbin\ (Figure~\ref{fig:mass-ratio}). We calculate the mass ratio
and its uncertainty from the range of model-inferred ages and the
luminosity ratio. Because the ratio of \Lbol\ is known to an
uncertainty of 0.008~dex, theoretical models make very precise
predictions of the mass ratio. The Tucson models give
$q$~=~0.962$\pm$0.003, and the Lyon models give $q$~=~0.948$\pm$0.005.
These model-inferred mass ratios are formally 2.4$\sigma$ discrepant,
but the resulting individual masses of \hdbin\ are completely
consistent because the 2\% error in the total mass dominates over the
0.3--0.5\% error in the mass ratio (Table~\ref{tbl:model}).  If we
were instead to use the age of \hdprim\ (\hdage) and luminosity ratio
of \hdbin\ to compute the mass ratio from the Tucson (Lyon)
evolutionary models, the result would differ by 1.2$^{+0.7}_{-0.5}$\%
(1.1$^{+0.9}_{-0.5}$\%).  Again, the error in the total mass dominates
over such small differences in mass ratio.

In principle, the mass ratio of \hdbin\ can be measured directly by
future resolved observations of the radial velocities of the two
components ($\Delta{v_{max}}$~=~6.4~\kms).  Such measurements will be
extremely valuable as they will test the model-predicted mass ratios,
which could harbor systematic errors.  However, given the attainable
radial velocity precision for L~dwarfs
\citep[0.1--0.3~\kms;][]{2007ApJ...666.1198B}, a future direct
measurement of the mass ratio to 2--5\% is unlikely to be precise
enough to discriminate between the two sets of evolutionary models, as
0.6\% errors in the mass ratio are needed to discriminate between the
model-inferred mass ratios at 90\% confidence level.

The individual masses of \hdbin\ inferred from theoretical models are
well below the substellar boundary of 0.072~\Msun\
\citep{2000ARA&A..38..337C}.  The Tucson models give masses of $M_{\rm
  B}$~=~0.0554$^{+0.0012}_{-0.0013}$~\Msun\ and $M_{\rm
  C}$~=~0.0532$^{+0.0012}_{-0.0011}$~\Msun, and the Lyon models give
$M_{\rm B}$~=~0.0558$^{+0.0012}_{-0.0012}$~\Msun\ and $M_{\rm
  C}$~=~0.0528$^{+0.0012}_{-0.0012}$~\Msun.  An extremely implausible
mass ratio of $\lesssim$0.5 would be required for HD~130948B to be a
star at the bottom of the main-sequence.  This scenario, as well as
the possibility that one component is an unresolved double, is
incompatible with the near-unity flux ratio and very similar
optical--near-infrared colors of the two components.  Thus, both
components of \hdbin\ are bona fide brown dwarfs.

\subsection{Luminosity Evolution \label{sec:lbol-evol}}

\begin{figure} 
\includegraphics[height=6.5in,angle=90]{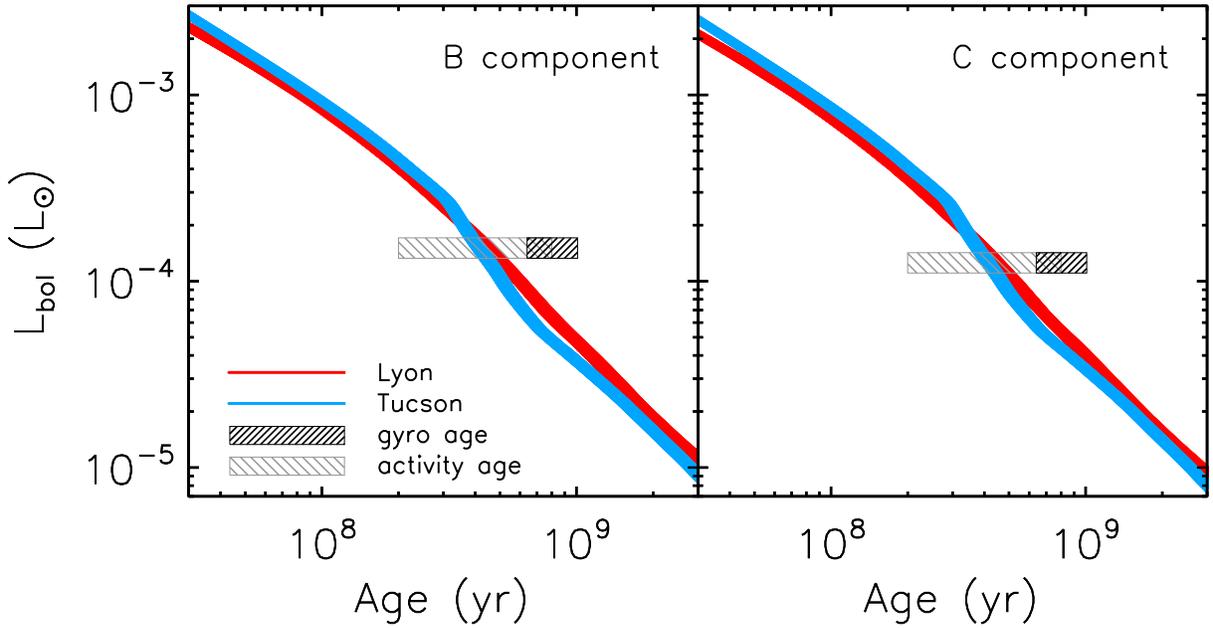}

\caption{ \normalsize Isomass lines for evolutionary models with the
  individual masses of \hdbin.  The colored line thicknesses encompass
  the 1$\sigma$ errors in the individual masses.  The hatched boxes
  indicate the constraints from the measured luminosities of
  HD~130948B and C and the age of \hdprim\ using gyrochronology and
  chromospheric activity.  The gyro age (\hdage) is inconsistent with
  the evolutionary models, implying that the models underpredict the
  luminosities of \hdbin\ by a factor of
  $\approx$2--3$\times$. \label{fig:lbol-age}}

\end{figure}

With well-determined masses and luminosities for \hdbin\ and an
independent age estimate from the primary star, we are able to
directly test one of the most fundamental predictions of substellar
theoretical models: the evolution of luminosity over
time. Figure~\ref{fig:lbol-age} shows the evolutionary model tracks
for objects with the individual masses of \hdbin\ compared to the
observations.  The Tucson and Lyon evolutionary models agree very well
with each other, which is one reason why they have become trusted to
estimate the bulk properties of brown dwarfs.  However, both sets of
models seem to disagree with the data.

Given the estimated age of \hdage, the Lyon models underpredict the
luminosity of both components of \hdbin\ by a factor of 2.3
(1.6--3.4$\times$, 1$\sigma$), and the Tucson models underpredict the
luminosities by a factor of 3.0 (2.1--4.3$\times$).  There are two
possible sources for the observed discrepancy in luminosity evolution.
The model radii could be at fault, in which case they would have to be
underpredicted by 20--45\% (30--50\%) by the Lyon (Tucson) models to
resolve the entire discrepancy.  Alternatively, the models may
correctly predict the radii of brown dwarfs but underpredict their
energy output.  As we will see in \S \ref{sec:teff-hr}, if the
evolutionary models indeed underpredict the luminosities of \hdbin\,
this has important implications for effective temperatures derived
from atmospheric models.

The severe disagreement between model-predicted and observed
luminosities is surprising, and we caution that it could be
ameliorated by a younger estimated age.  A younger age could be
accommodated by most of the age indicators (Table~\ref{tbl:age}), and
given the challenges in estimating ages of field main-sequence stars
this possibility cannot be ignored.  For example, the gyro age is not
beyond reproach (see \S \ref{sec:age-summ}), and the less precise
activity age (0.5$\pm$0.3~Gyr) allows for better agreement between the
models and the data.  However, as discussed in \S \ref{sec:age-summ}
our lower limit on the gyro age is expected to be robust, and \hdprim\
generally appears to be very consistent with the age of the Hyades.
At this age, both components of \hdbin\ would still be more luminous
than predicted by evolutionary models, though the discrepancy would be
on the lower end of the ranges given above.  We note that
\citet{gl802b-ireland} has observed a similar effect for GJ~802B, a
substellar ($M$~=~0.063$\pm$0.005~\Msun) companion to a kinematically
old star ($\sim$10~Gyr), for which the evolutionary models predict an
age of $\sim$2~Gyr given its mass and luminosity.

\subsection{Color-Magnitude Diagrams \label{sec:cmd}}

The Lyon evolutionary models provide predictions of the fluxes in
various observational bandpasses for each model mass and age.
Figure~\ref{fig:jhk-age} shows how the predicted evolution of the
near-infrared flux of \hdbin\ compares to the observations.  Given the
age of \hdprim\ (\hdage), the evolutionary models underpredict the
flux in every bandpass.  This is simply a reflection of the fact that
Lyon evolutionary models underpredict the luminosity for both
components of \hdbin\ for the age of \hdprim.  However, if ages were
actually inferred from the Lyon models using the observed $J$-, $H$-,
and $K$-band photometry, Figure~\ref{fig:jhk-age} shows these ages
would not be self-consistent.  Thus model-predicted near-infrared
magnitudes are internally inconsistent with the data.

In Figure~\ref{fig:jhk-cmd} we show the predicted $JHK$ colors of both
components of \hdbin\ compared to the observations.  The $J-K$ and
$H-K$ colors are both significantly discrepant for any assumed age,
while the $J-H$ colors seem to agree with the models at an age younger
than we have estimated for \hdprim.  (We do not place significant
weight on this agreement as the $J-H$ color measurement has the
largest uncertainty.)  The general disagreement between models and
data we observe on the color-magnitude diagram is not surprising since
it is well-known that theoretical models do not reproduce the
near-infrared colors of field L and T~dwarfs very well
\citep[e.g.,][]{2004AJ....127.3553K,2005astro.ph..9066B}.

It is interesting to note that if we were to infer the ages and masses
of the components of \hdbin\ from the model $J-K$ or $H-K$
color-magnitude diagrams, we would derive masses that are
$\approx$20--30\% smaller and ages $\approx$2$\times$ younger than
observed.  Thus, masses and ages inferred for L~dwarfs from
evolutionary models and near-infrared photometry should be treated
with caution.

\subsection{Temperatures and Surface Gravities \label{sec:teff-logg}}

Without radii measurements for \hdbin, we must rely on evolutionary
models to derive effective temperatures and surface gravities.  We
have several independent measurements of the fundamental properties of
\hdbin\ at our disposal to use with models to find \Teff\ and \logg:
(1) the total mass of the system; (2) the individual luminosities of
the two components; (3) the luminosity ratio; and (4) an independent
age estimate from the primary \hdprim\ (\hdage).  We use the total
mass (1) and individual luminosities (2) to derive \Teff\ and \logg\
from the evolutionary models (Table~\ref{tbl:model}).  This is
conceptually equivalent to using the individual masses (\S
\ref{sec:qratio}) and model-inferred age (\S \ref{sec:modelage}) to
derive \Teff\ and \logg.  Unlike previous sections, it now matters
significantly whether we use the model-inferred ages or the
independent age from \hdprim\ (\hdage) to derive the quantities of
interest.  We will describe the resulting differences later in this
section.

The Lyon models give effective temperatures for B and C of
1990$\pm$50~K and 1900$\pm$50~K, while the Tucson models give
systematically hotter but formally consistent temperatures of
2040$\pm$50~K and 1950$\pm$50~K. Since brown dwarfs cool over time, it
is essentially the small range of model-inferred ages which allows the
effective temperature to be predicted to a precision of 50~K.  The
Lyon models give surface gravities for B and C of
\logg~=~5.143$\pm$0.019 and 5.122$\pm$0.019 (cgs), while the Tucson
models give systematically higher and formally inconsistent gravities
of \logg~=~5.196$\pm$0.017 and 5.183$\pm$0.017 (cgs). The precision in
model-inferred surface gravity is driven by the precision in the
measured total mass and near-unity mass ratio, since the radii of
brown dwarfs remain nearly constant after 0.3~Gyr.  Thus, the
difference between the two sets of model-inferred surface gravities
arises from small differences ($\approx$6\%) in their predictions for
the radii (Table~\ref{tbl:model}).

\subsubsection{Comparison to Field L~Dwarfs \label{sec:teff-field}}

The effective temperatures we derive from evolutionary models can be
compared to those which have been determined for other objects of
\hdbin's spectral type.  L3--L5 dwarfs in the field with \Lbol\
measurements have estimated effective temperatures of
1650--2050~K. These estimates utilize the nearly flat mass-radius
relationship predicted by theoretical models for brown dwarfs,
adopting either a typical age \citep[3~Gyr;][]{gol04} or radius
\citep[0.90$\pm$0.15~\Rsun;][]{2004AJ....127.2948V} for field objects.
It has been suggested that the ages of field objects are overestimated
\citep[e.g.,][]{2006ApJ...651.1166M,2008arXiv0807.0238L}; however the
broad range of \Teff\ estimated for field L3--L5 dwarfs is consistent
with our model-inferred effective temperatures.  Since both estimates
are based on evolutionary models, this only means that field L3--L5
dwarfs from previous studies encompass objects of the same mass/age as
\hdbin.

\subsubsection{Comparison to Atmospheric Models \label{sec:teff-atm}}

For a more interesting comparison, we consider effective temperatures
derived from spectral synthesis using state-of-the-art atmospheric
models.  \citet{2008ApJ...678.1372C} have performed the most thorough
spectral synthesis analysis of L and T dwarfs to date, and four of the
objects in their study have near-infrared spectral types of L3--L5
(including 2MASS~J0036+1821, the best matching spectral template from
\citealp{2002ApJ...567L..59G}).  Their fits to the 0.95--14.5~\micron\
spectra of L3--L5 dwarfs yield effective temperatures of 1700--1800~K,
which are significantly cooler (150--300~K) than the evolutionary
model-inferred temperatures for \hdbin.  Adopting 1750$\pm$100~K as
the atmospheric model \Teff\ for the B and C components, we find the
significance of the discrepancy with the Lyon (Tucson) model-inferred
temperature is 2.6$\sigma$ (2.1$\sigma$) for the B component and
1.8$\sigma$ (1.3$\sigma$) for C.  This discrepancy cannot simply be
due to our adopted mass ratio because, for example, if the mass ratio
were tuned so that the C component was lower mass (thus cooler) the B
component would become more massive (thus hotter) and even more
discrepant with the effective temperatures from spectral synthesis.

\citet{2008ApJ...678.1372C} also derive surface gravities for the four
L3--L5 dwarfs in their study.  They do so both by direct model fitting
(4.5--5.5) and by using evolutionary sequences
(4.9--5.5),\footnote{\citet{2008ApJ...678.1372C} fit for \Teff, \logg,
  and a normalization constant $(R/d)^2$.  Thus, for the three of the
  four L3--L5 dwarfs with parallax measurements, they also have radius
  estimates.  By using evolutionary models, \Teff\ and $R$ uniquely
  determine \logg. } and these ranges are consistent with our
model-inferred values of \logg\ for \hdbin.

\subsubsection{Using the Age of \hdprim \label{sec:teff-primage}}

The effective temperatures we derive for \hdbin\ from the Lyon and
Tucson evolutionary models depend greatly on whether we use the
model-inferred ages (0.45$^{+0.05}_{-0.04}$~Gyr and
0.41$^{+0.04}_{-0.03}$~Gyr, respectively) or the independent age
estimate from \hdprim\ (\hdage).  Using the age of \hdprim, combined
with the masses of \hdbin, the Lyon (Tucson) models give effective
temperatures of 1670$^{+120}_{-110}$~K and 1600$^{+100}_{-110}$~K
(1590$^{+110}_{-100}$~K and 1550$^{+90}_{-90}$~K) for the B and C
components, respectively.  These temperatures are 300--450~K cooler
than those derived using the model-inferred age because the masses are
the same, but the age used is significantly older.  This disagreement
is just a restatement that the model-inferred age and gyro age do not
agree.

The values of \Teff\ derived using the age of \hdprim\ also disagree
with those determined for L3--L5 dwarfs by spectral synthesis fitting
\citep{2008ApJ...678.1372C}.  They are 100--150~K cooler than the
atmospheric model temperatures.

We note that all these different sets of effective temperatures imply
different radii for \hdbin, since the luminosities of the two
components are well-determined.  We consider the interplay between
\Lbol\ and \Teff\ in more detail in \S \ref{sec:teff-hr}.

\subsubsection{$\Delta$\Teff\ Compared to $\Delta$SpT \label{sec:teff-dteff}}

Despite the fact that the components of \hdbin\ are nearly twins in
mass and spectral type, evolutionary models predict rather large
differences in the effective temperatures of the two components
(90$\pm$7~K and 85$\pm$7~K for the Tucson and Lyon models,
respectively).\footnote{Note that the error on the temperature
  difference is determined by the precision in the luminosity ratio,
  so it is much smaller than the 50~K uncertainties in individual
  values of \Teff.}  This is computed directly from evolutionary
models using the model-inferred age and measured luminosity ratio.
Such a large effective temperature difference may be discernable in
future spectral synthesis modeling of the resolved SED of \hdbin\
using integral field spectroscopy, which is not subject to the same
difficulties as AO-fed slit spectroscopy.  

We could estimate a 90~K difference in \Teff\ even without using
models given the luminosity ratio and an assumption that the radii of
the two components are roughly equal (i.e.,
$\Delta\Teff/\Teff$~$\propto$~$(\Delta\Lbol/\Lbol)^{1/4}$).  However,
this value of $\Delta$\Teff\ is perhaps somewhat surprising since
\citet{2002ApJ...567L..59G} found that the two components are nearly
twins in spectral type, and any AO-related modifications to the
continuum shape should affect both components equally.  According to
the SpT--\Teff\ relation of \citet{gol04}, model-inferred temperatures
of \hdbin\ give a difference in spectral type of $\approx$1 subtype.
This lack of an apparent change in spectral type with \Teff\ may be
indicative of other atmospheric processes, such as condensate cloud
formation and sedimentation, playing a role that is at least as
important as temperature in shaping the emergent spectra of mid-L
dwarfs.  In other words, this is suggestive that spectral type may not
have a one-to-one correspondence with effective temperature for mid-L
dwarfs \citep[see][]{kirk05}.

\subsection{H-R Diagram \label{sec:teff-hr}}

\begin{figure} 
\includegraphics[height=6.5in,angle=90]{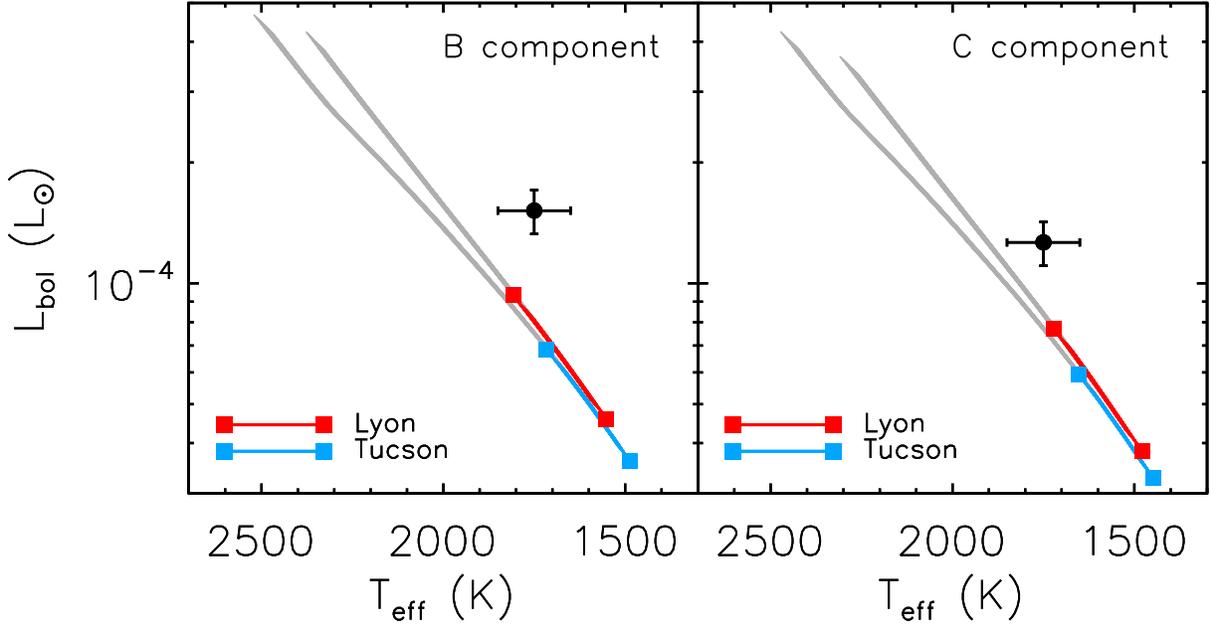}

\caption{ \normalsize The Hertzsprung-Russell diagram showing isomass
  lines from evolutionary models corresponding to the individual
  masses of \hdbin.  The plotted line thicknesses of these tracks
  encompass the 1$\sigma$ errors in the individual masses.  The red
  and blue colored regions correspond to the gyro age for \hdprim\
  (\hdage).  The gray shaded regions correspond to the less precise
  chromospheric activity age (0.5$\pm$0.3~Gyr).  The Lyon and Tucson
  evolutionary models are nearly indistinguishable in their predicted
  \Teff\ and \Lbol\ at these ages.  The effective temperature
  determined for field L3--L5 dwarfs from spectral synthesis
  (1750$\pm$100~K) is shown as a filled circle at the measured
  luminosities of HD~130948B and C.  The numerous discrepancies
  between the models and the data seen here are discussed in \S
  \ref{sec:teff-hr}.  \label{fig:hr-diagram}}

\end{figure}

\begin{figure} 
\includegraphics[height=5.5in,angle=90]{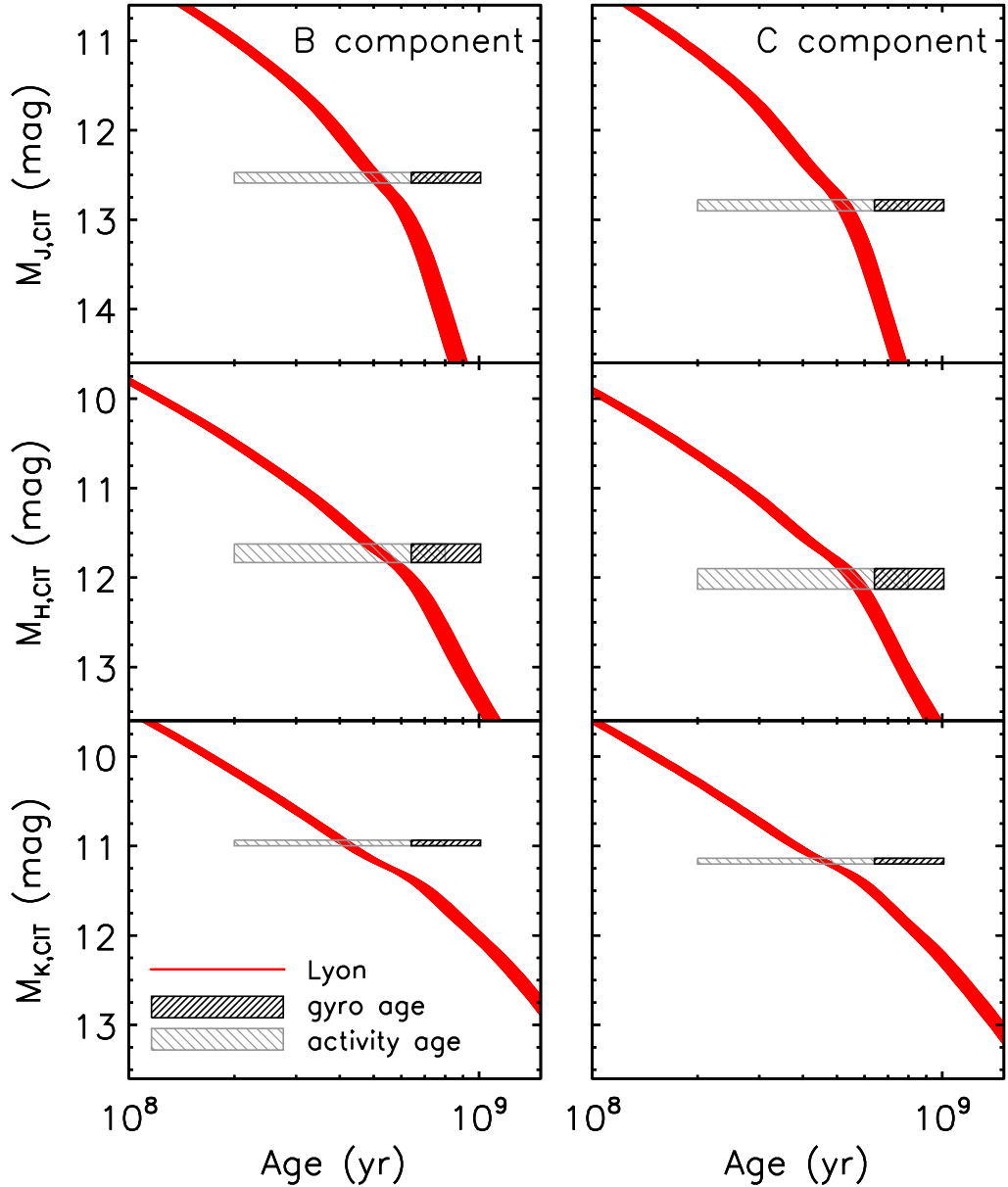}

\caption{ \normalsize Isomass lines showing the evolution of $J$-,
  $H$-, and $K$-band absolute magnitudes for Lyon models with the
  individual masses of \hdbin. The colored line thicknesses encompass
  the 1$\sigma$ errors in the individual masses.  The hatched boxes
  indicate the constraints from the age of \hdprim\ and the measured
  photometry of HD~130948B and C.  The photometry shown here is on the
  CIT system, where we have converted our measured photometry of
  \hdbin\ from the MKO system using the relations of
  \citet{2004PASP..116....9S}.  The gyro age (\hdage) is inconsistent
  with the predicted fluxes, which is a reflection of the
  underpredicted bolometric luminosities of \hdbin\
  (Figure~\ref{fig:lbol-age}).  The model tracks intersect the
  measured photometry at different ages for different filters, which
  indicates inconsistencies in the model-predicted near-infrared
  colors (see Figure~\ref{fig:jhk-cmd}).  \label{fig:jhk-age}}

\end{figure}

\begin{figure} 
\includegraphics[height=6.5in,angle=90]{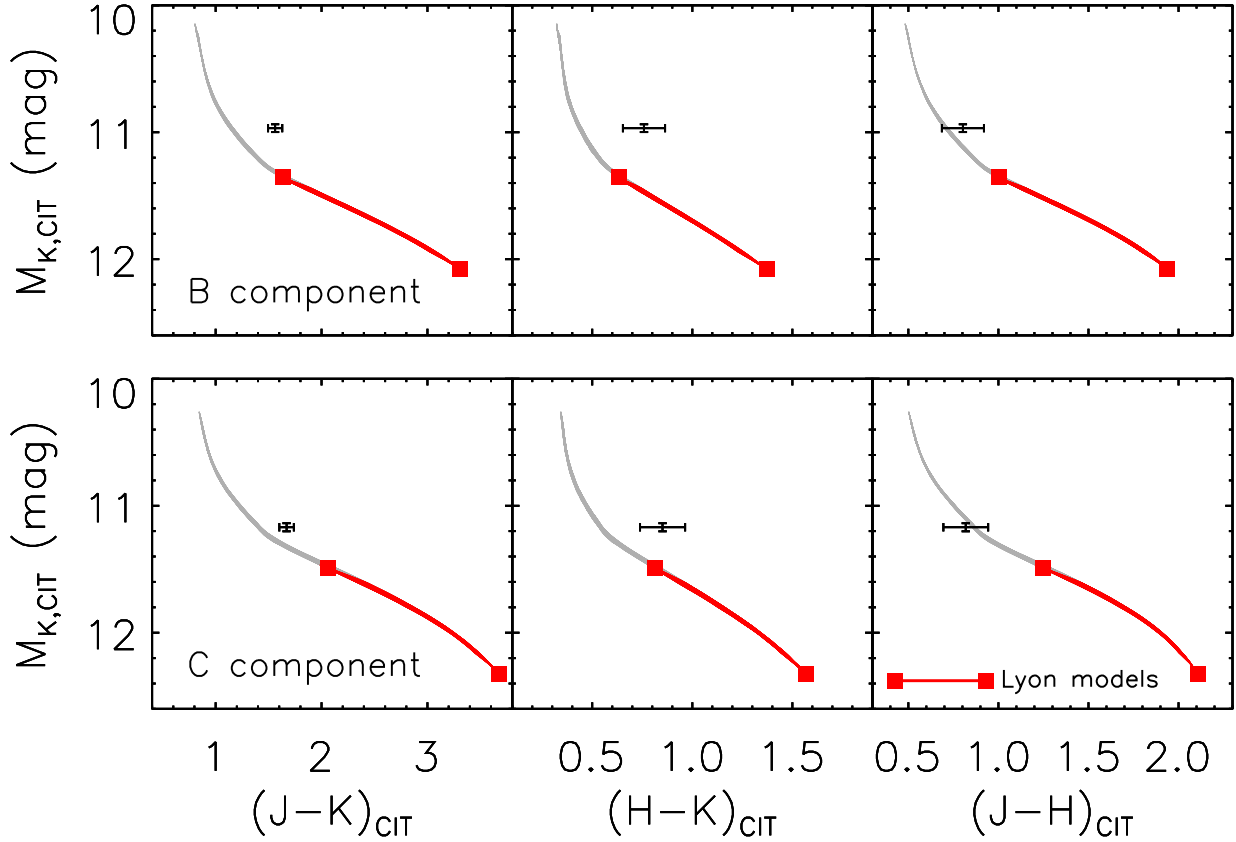}

\caption{ \normalsize Color-magnitude diagrams showing the measured
  properties of \hdbin\ compared to Lyon evolutionary tracks for the
  masses of B and C.  The plotted line thicknesses encompass the
  1$\sigma$ errors in the individual masses.  The red colored regions
  correspond to the gyro age for \hdprim\ (\hdage).  The gray shaded
  regions correspond to the less precise activity age
  (0.5$\pm$0.3~Gyr).  In general, the measured colors are discrepant
  from that predicted by evolutionary models.  Thus, evolutionary
  models will generally not yield accurate mass and/or age estimates
  for field L~dwarfs from techniques using color-magnitude diagrams
  alone.  \label{fig:jhk-cmd}}

\end{figure}

In the previous section, we have determined various effective
temperatures for \hdbin\ from evolutionary and atmospheric models
given the available observational constraints.  We now combine these
temperatures with the observed luminosities of \hdbin\ to place both
components on the Hertzsprung-Russell diagram
(Figure~\ref{fig:hr-diagram}).  The H-R diagram shows substantial
discrepancies between evolutionary models, atmospheric models, and the
observations.  Namely, both components of \hdbin\ are more luminous
than predicted by evolutionary models for objects of their masses and
effective temperatures, where \Teff\ is independently adopted from
atmospheric models.  Alternatively, the discrepancy in
Figure~\ref{fig:hr-diagram} may be stated as both components of
\hdbin\ being cooler than predicted by evolutionary models given their
masses and luminosities.

The age is the least precisely determined fundamental parameter for
\hdbin\ ($\approx$25\%), since the masses and luminosities are
accurate to 2\% and 13\%, respectively.  Thus, we can consider whether
changing the adopted age of the system can resolve the discrepancies
between the data and models revealed in the H-R diagram.  There are 3
plausible scenarios:

\begin{itemize}

\item If our preferred \hdage\ age for \hdprim\ is correct, then
  evolutionary models underpredict the luminosities of \hdbin\ by a
  factor of $\approx$2--3 (\S \ref{sec:lbol-evol}), and atmospheric
  models predict temperatures that are 100-150~K warmer than
  evolutionary models (\S \ref{sec:teff-primage}).

\item If the system has a slightly younger age of $\approx$0.6~Gyr,
  then the effective temperatures predicted by evolutionary and
  atmospheric models would agree.  However, the evolutionary models
  would still underpredict the luminosities of \hdbin\ by a factor of
  $\approx$1.5--2.  The age in this scenario is consistent with all
  available age indicators (see \S \ref{sec:age-summ}).

\item If the system is as young as $\approx$0.4~Gyr, then the
  evolutionary models would predict the correct luminosities.
  However, atmospheric models would then indicate temperatures
  150--300~K cooler than predicted by evolutionary models (\S
  \ref{sec:teff-atm}).  In other words, if the actual age was
  consistent with the model-inferred ages of
  0.45$^{+0.05}_{-0.04}$~Gyr and 0.41$^{+0.04}_{-0.03}$~Gyr, then the
  predicted and observed luminosities would agree by construction.
  However, the age in this scenario is significantly discrepant
  ($>$2$\sigma$) with the gyro age.

\end{itemize}

{\em Thus, no scenario exists in which both evolutionary and
  atmospheric models agree with the data.}  In \S \ref{sec:lbol-evol}
we have already discussed the possible sources of systematic errors in
evolutionary models.  A number of causes might be responsible for
systematic errors in the atmospheric models, including insufficient
treatment of dust in the photosphere, incomplete line lists, and/or a
metallicity bias in the \citet{2008ApJ...678.1372C} sample.

A refined age estimate for \hdprim\ will be essential in discerning
the source of disagreement between models and data.  Another important
step will be obtaining resolved spectroscopy of \hdbin\ suitable for
direct atmospheric model fitting, which will reduce the uncertainties
in \Teff.  The published AO-fed slit spectroscopy is not sufficient
for this task (see \S \ref{sec:spt}), but ground-based AO integral
field spectroscopy in the near-infrared and space-borne spectroscopy
in the optical would be ideal once \hdbin\ is resolvable again in
2010.

\subsection{Lithium Depletion \label{sec:lithium}}

Structural models used in the prediction of brown dwarf cooling
sequences make direct predictions of the amount of lithium depletion
that has occurred for an object of a given mass and age. Because brown
dwarfs are fully convective objects, the depletion of lithium
throughout the entire object is readily detectable from observations
of photospheric absorption lines, the strongest of which is the
doublet at 6708~\AA. Both the Tucson and Lyon models predict that
objects less massive than $\approx$0.06~\Msun\ never reach internal
temperatures high enough to destroy significant amounts of their
primordial lithium \citep{bur01,2000ApJ...542..464C}. Since higher
mass objects deplete lithium faster, \citet{2005astro.ph..8082L}
proposed that ultracool binary systems caught at just the right point
in their evolution would enable a very precise age estimate if the
less massive component was found to possess lithium and the more
massive component was lithium-depleted.  Therefore, L dwarf binaries
displaying lithium in their unresolved spectra \citep[6 are known to
date; see list in][]{2005astro.ph..8082L} are potentially powerful
systems for constraining theoretical models of brown dwarfs, since
they are amenable to this ``binary lithium test''.

\begin{figure}
\centerline{\includegraphics[width=6.0in,angle=0]{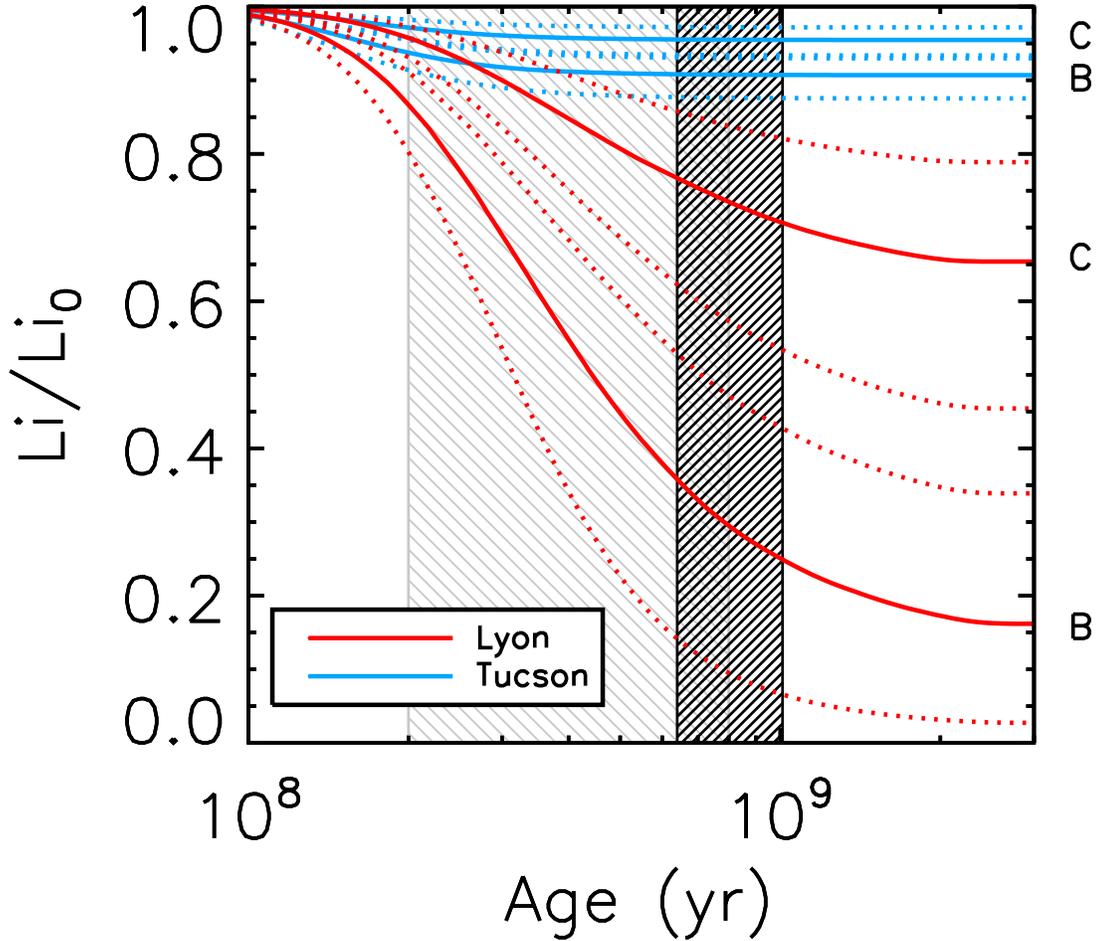}}

\caption{ \normalsize Lithium depletion as a function of age as
  predicted by evolutionary models.  The solid lines correspond to the
  individual masses of the B and C components of \hdbin\ (C has a
  higher lithium fraction).  These lines are bracketed by dotted lines
  that correspond to the 1$\sigma$ uncertainties in the individual
  masses.  The ordinate is the fraction of primordial lithium
  remaining.  The hatched black (gray) box indicates the constraint
  from the age of \hdprim\ estimated from gyrochronology
  (chromospheric activity).  The Lyon models predict that all objects
  more massive than $\gtrsim$0.055~\Msun\ eventually deplete most of
  their primoridal lithium, while the Tucson models predict that this
  occurs only for objects more massive than $\gtrsim$0.062~\Msun.  For
  the Lyon models, the components of \hdbin\ straddle the
  lithium-burning limit, while for the Tucson models neither component
  is expected to have depleted a significant amount of lithium.  Thus,
  the age and individual masses of \hdbin\ are ideal for
  discriminating between these two sets of models with resolved
  optical spectroscopy designed to detect lithium absorption at
  6708~\AA.  This would provide the only direct empirical constraint
  to date on the theoretical timescale and mass limit of
  lithium-burning in brown dwarfs, which are the basis for the
  ``cluster lithium test'' \citep[e.g.,][]{1998bdep.conf..394B} and
  the proposed ``binary lithium test''
  \citep{2005astro.ph..8082L}.  \label{fig:lithium}}

\end{figure}

There is no optical spectroscopy (resolved or unresolved) available
for \hdbin\ to assess the presence of lithium in the system. However,
the individual masses of \hdbin\ are very close to, perhaps
straddling, the theoretical lithium-burning limit
(Figure~\ref{fig:lithium}). We have determined the model-predicted
lithium abundance for each component from the model-inferred age and
individual masses. The Tucson models predict very little lithium
depletion for both objects, with the C component having only slightly
higher lithium abundance than the B component by a factor of
1.049$^{+0.011}_{-0.014}$.  (The error in the relative lithium
abundance is dominated by the uncertainty in the model-inferred age.)
On the other hand, the Lyon models predict that B is massive enough
that it has depleted most of its primordial lithium
(Li/Li$_0$~=~0.50$^{+0.18}_{-0.23}$) while C has retained most of its
lithium (Li/Li$_0$~=~0.83$^{+0.08}_{-0.13}$). In fact, lithium burning
occurs so quickly that even over the small range of Lyon
model-inferred ages (0.45$^{+0.05}_{-0.04}$ Gyr) the amount of
relative lithium depletion between B and C is quite uncertain (i.e., C
is predicted to be richer than B by a factor of 1.6$^{+1.0}_{-0.3}$).

The individual masses of \hdbin\ are such that the presence or absence
of lithium in their resolved spectra would provide significant
discrimation between the Tucson and Lyon models. Future resolved
optical spectroscopy of \hdbin\ will provide a very sensitive, direct
test of the lithium-burning limit for brown dwarfs. Given the
independent age constraint from \hdprim, the theoretical timescale for
lithium burning can also be directly tested if one or both components
of \hdbin\ show evidence of lithium depletion. Such direct tests of
theoretical predictions for lithium burning would provide the only
empirical calibration to date of often used theoretical predictions of
lithium burning in brown dwarfs.  These predictions have provided the
basis of the ``binary lithium test'', as well as the more widely known
``cluster lithium test'' \citep[e.g.,][]{1998bdep.conf..394B} used to
identify substellar objects among associations of a known age.


\section{Conclusions}

We have determined the orbit of the young L4+L4 binary \hdbin\ using
relative astrometry of the system spanning 7 years of its 10-year
orbital period.  The astrometric measurements and their uncertainties
were extensively tested through Monte Carlo simulations.  The fitted
orbital parameters and revised \Hipparcos\ parallax give a total
dynamical mass of 0.109$\pm$0.002~\Msun.  The precision in mass is
2\%, with nearly equal contributions to the uncertainty from the 1.7\%
error in the best-fit orbit and the 1.3\% error in mass from the
\Hipparcos\ parallax error.  For any plausible mass ratio, both
components of \hdbin\ are unambiguously substellar.  \hdbin\ has the
most precise mass determination for a brown dwarf binary to date.

The primary star \hdprim\ offers an independent constraint on the age
of the system from various indicators: rotation, chromospheric
activity, isochrone fitting, X-ray emission, and lithium depletion.
The ensemble of all available age indicators is consistent with an age
for \hdprim\ similar the Hyades (625~Myr).  For example, its rotation
period is inconsistent with ages much younger than the Hyades and its
chromospheric activity is inconsistent with ages much older than the
Hyades.  Our preferred age estimate is \hdage, derived from the
gyrochronology formalism of \citet{2007ApJ...669.1167B} and
\citet{mam08-ages}.

With a measured mass, luminosity, and age, \hdbin\ provides the first
direct test of the luminosity evolution predicted by theoretical
models for substellar field dwarfs.  Both the Tucson models
\citep{1997ApJ...491..856B} and Lyon models
\citep[DUSTY;][]{2000ApJ...542..464C} underpredict the luminosities of
HD~130948B and C given their masses and age.  The discrepancy is quite
large, about a factor of 2 for the Lyon models and a factor of 3 for
the Tucson models.  In order to explain this discrepancy entirely,
model radii would have to be underpredicted by 30--40\%.  The age of
\hdprim\ would need to be $\approx$0.4~Gyr younger than we have
estimated in order to resolve this discrepancy.  This is inconsistent
with the preferred gyro age but can be accommodated by other age
indicators; a more refined age estimate for \hdprim\ is critically
needed.

Since the mass of \hdbin\ is more precisely determined than its age,
we have used the mass with the individual bolometric luminosities to
infer all other properties (age, \Teff, etc.) from evolutionary
models.  We use a Monte Carlo approach to compute model-inferred
quantities, and we are careful to account for covariance between the
observational errors, the most notable of which is the correlation of
the luminosities of the two components through their measured flux
ratio.  Because we use mass and \Lbol\ to derive model-inferred
properties, any potential systematic errors in luminosity evolution
will be reflected in the model-inferred quantities.  For example, the
very precise model-inferred ages for \hdbin\
(0.41$^{+0.04}_{-0.03}$~Gyr from Tucson models;
0.45$^{+0.05}_{-0.04}$~Gyr from Lyon models) are self-consistent, but
they are inconsistent with the independent age estimate for \hdprim\
(\hdage).

Lacking measured radii for \hdbin, we have used evolutionary models to
derive effective temperatures.  Given the mass and luminosity of each
component, evolutionary models predict effective temperatures of
$\approx$1900--2000~K.  Alternatively, given the mass of each
component and age of the primary star, evolutionary models predict
effective temperatures of $\approx$1600--1700~K.  (The disagreement
between these two temperature ranges is just a reflection of the
systematic errors in luminosity evolution.)  Spectral synthesis using
atmospheric models gives temperatures of 1700--1800~K for objects of
similar spectral type to \hdbin\
\citep[L3--L5;][]{2008ApJ...678.1372C}.  Using evolutionary models and
the measured luminosity ratio gives $\Delta$\Teff~=~90~K.  Resolved
spectroscopy of HD~130948B and C has previously shown that they have
indistinguishable spectral types, so this rather large temperature
difference may indicate that spectral type does not hold a one-to-one
correspondence with \Teff\ mid-L~dwarfs, even for two coeval objects.
Better spectral types for the two components of \hdbin\ are needed to
address this apparent discrepancy.

Comparing the different effective temperature determinations for
\hdbin\ on the H-R diagram shows that the evolutionary models,
atmospheric models, and observational data cannot be simultaneously
brought into consistency with each other, regardless of the adopted
age of the system.  Thus, systematic errors in some combination of the
atmospheric and/or evolutionary models are needed to explain the
observed discrepancy.  The best current age estimate indicates that
both evolutionary and atmospheric models harbor systematic errors.
Further evaluation of the disagreement between models and the data
requires a refined age estimate for \hdprim.  Resolved multi-band
spectroscopy of \hdbin\ is also needed to reduce the uncertainties in
the atmospheric model effective temperatures by direct spectral
synthesis fitting.

We also find large discrepancies when comparing the observed
near-infrared colors of \hdbin\ to the Lyon models.  This suggests
that using color-magnitude diagrams to infer the properties of field
L~dwarfs from evolutionary models will lead to large errors in the
resulting quantities (e.g., mass and/or age).  For example, if we
inferred the ages and masses of the components of \hdbin\ from the
model $J-K$ or $H-K$ color-magnitude diagrams, we would derive masses
that are $\approx$20--30\% smaller than observed and ages
$\approx$2$\times$ younger than the age of the primary star (\hdage).

One novel aspect of using \hdbin\ to constrain theoretical models is
the application of the ``binary lithium test'', originally proposed by
\citet{2005astro.ph..8082L}.  This is made possible by the fortuitous
circumstance that the components of \hdbin\ are very near the mass
limit for lithium burning.  As a consequence, the Lyon and Tucson
evolutionary models, which are almost indistinguishable in their
predictions of substellar bulk properties, give very different
predictions for the amount of primordial lithium remaining in the B
and C components.  Thus, resolved optical spectroscopy to detect the
lithium doublet at 6708~\AA\ would provide a very discriminating test
of the evolutionary models.  Such a constraint is significant in that
it directly tests the properties of fully convective substellar
interiors (e.g., the core temperature) and/or the lithium reaction
rates.  \hdbin\ is the only system currently known for which such an
empirical calibration of lithium burning is possible.

Substellar theoretical models are in sore need of empirical validation
as they have been employed for more than a decade to interpret
observations of field dwarfs.  Given the independent constraints on
the age and composition provided by a stellar companion, dynamical
mass measurements for triple systems like HD~130948ABC provide the
most challenging tests of substellar theoretical models.  However,
substellar companions to stars are quite rare
\citep[$\approx$1$\pm$1\%, e.g.,][]{2001AJ....121.2189O,
  2004AJ....127.2871M, 2005AJ....130.1845L, 2007ApJS..173..143B,
  2007ApJ...670.1367L}, and even more rare are substellar binary
companions that yield dynamical mass measurements in a reasonable time
frame.  When the stellar companion is a bright star like \hdprim, a
wealth of additional information is available, the most important of
which is a very precise \Hipparcos\ distance measurement since this is
the limiting factor in the precision of the dynamical mass.  Stars
bright enough to enable seismological measurements can yield the most
stringent (10--20\%) age determinations possible
\citep[e.g.,][]{2005A&A...434.1085C,2008ApJ...673.1093B}.  Thus,
\hdbin\ represents a rare class of benchmark systems for which the
most precise mass and age determinations are possible.

Our observations of \hdbin\ indicate that substellar models currently
harbor significant systematic errors.  The potential underestimation
of \Lbol\ by evolutionary models has far-reaching implications.  For
example, such models have been used to determine the low-mass end of
the intial mass function and to predict the radii of extrasolar
planets.  Obtaining measurements for more systems like \hdbin\ over a
broad range of mass, luminosity, and age will be critical in
understanding and resolving the discrepancies that have been revealed
between observations and theoretical models.


\acknowledgments

We thank Colin Cox and John Krist for their scientific foresight in
selecting \hdprim\ as the target for \HST\ engineering and calibration
observations.  We also thank John Krist for assistance with the
TinyTim software.  We gratefully acknowledge the Keck AO team for
their exceptional efforts in bringing the AO system to fruition.  It
is a pleasure to thank Antonin Bouchez, David LeMignant, Marcos van
Dam, Randy Campbell, Al Conrad, Jim Lyke, Hien Tran, Joel Aycock,
Julie Rivera, Cindy Wilburn, Jason McIlroy, and Gary Punawai and the
Keck Observatory staff for assistance with the observations.  We are
grateful to Brian Cameron for making available his NIRC2 distortion
solution, Lynne Hillenbrand for providing an advance copy of the
manuscript relating chromospheric activity and age, and Adam Burrows
and Isabelle Baraffe for providing finely gridded evolutionary models.
We have benefitted from discussions with Michael Cushing about
theoretical models, Brian Cameron about astrometry with NIRC2, Jay
Anderson about astrometry with \HST, Thierry Forveille about orbit
fitting, and Hai Fu about data analysis and plotting.  We also thank
Zahed Wahhaj for his help with the 2007 July observations and Michael
Cushing for providing us with the reduced IRTF/SpeX spectrum of
HD~130948A.
Our research has employed the 2MASS data products; NASA's
Astrophysical Data System; the SIMBAD database operated at CDS,
Strasbourg, France; and the M, L, and T~dwarf compendium housed at
DwarfArchives.org and maintained by Chris Gelino, Davy Kirkpatrick,
and Adam Burgasser \citep{2003IAUS..211..189K, 2004AAS...205.1113G}.
TJD and MCL acknowledge support for this work from NSF grant
AST-0507833, and MCL acknowledges support from an Alfred P. Sloan
Research Fellowship.
Finally, the authors wish to recognize and acknowledge the very
significant cultural role and reverence that the summit of Mauna Kea has
always had within the indigenous Hawaiian community.  We are most
fortunate to have the opportunity to conduct observations from this
mountain.

{\it Facilities:} \facility{Keck II Telescope (NGS AO, NIRC2)},
\facility{\HST/ACS}, \facility{Gemini-North Telescope}




\appendix

\section{Monte Carlo Simulations of \hdbin\ in \HST/ACS Coronagraph Images \label{app:hst}}

In order to robustly determine the systematic and random uncertainties
in our PSF-fitting measurements of \hdbin\ in the 2002 and 2005
\HST/ACS coronagraph data, we fit an array of simulated binary images
constructed from images of single stars.  No suitable single stars
were present in any archival coronagraph images taken in the same
filters as \hdbin\ ($F850LP$ and $FR914M$).  Therefore, we turned to
the much richer archive of $F814W$ coronagraph data, in which we found
numerous suitable single stars.  We selected three of the highest
$S/N$ stars in the archive, all of which come from a 2003 March 25 UT
observation of HD~163296 (GTO/ACS-9295, PI Ford).  These two
$\approx$2000 sec exposures were taken at two different roll angles,
providing optimal subtraction of the background light due to the
bright occulted star, and we used the second image (14:15~UT) to
subtract the background from the first (11:07~UT).  The locations of
the three stars we selected sample very different subpixel locations,
which is a potential source of systematic error for the fitting of the
slightly undersampled ACS PSF.  In the end, we found that no matter
which single star we used, the resulting astrometry did not change
significantly, so all results we quote from our Monte Carlo
simulations refer to the highest $S/N$ star.  This star was scaled
down by 3.3--5.7 mag to match the $S/N$ of the science data, depending
on the epoch and bandpass.

We created simulated binary images at the integer pixel separations
that most closely approximated \hdbin\ at each epoch and telescope
roll angle.  Subpixel-shifted binary images are impossible to
accurately create from one image of a single star because the ACS-HRC
PSF is slightly undersampled, inhibiting accurate interpolation to a
fraction of a pixel.  For the 2002 epoch, we used an equal number of
simulated binaries with integer separations of
$(\Delta{x},\Delta{y})=(-3, 3)$ and $(\Delta{x},\Delta{y})=(-2, 3)$ to
approximate \hdbin\ which had a measured separation of
$(\Delta{x},\Delta{y})\approx(-2.5, 3.0)$.  For the 2005 epoch, we
used simulated binaries at integer pixel separations of
$(\Delta{x},\Delta{y})=(-2, 1)$ and $(\Delta{x},\Delta{y})=(-1, 2)$.
Each corresponds to a different roll angle in the science data, for
which the actual best-fit separations were
$(\Delta{x},\Delta{y})\approx(-1.8, 1.2)$ and
$(\Delta{x},\Delta{y})\approx(-1.4, 1.8)$.

We scaled down the simulated binary images to match the peak counts of
the science data then added photon noise assuming a gain of
2.2~e$^-$/DN (from the \texttt{ATODGAIN} header keyword).  We used a
flux ratio of 0.25~mag, consistent with the flux ratio measured in the
2005 $F850LP$ data, when creating all simulated binary images.  Since
\hdbin\ itself is always located in the exact part of the residual
background in which we would like to inject our simulated binaries in
any given image, we instead injected them at a location
180\degs-symmetric to the location of \hdbin.  This is motivated by
the fact that the background light is visibly 180\degs-symmetric even
on scales as small as a few pixels.  (Note that this symmetry is not
so perfect that rotated self-subtraction is preferred over roll
subtraction for the removal of background light to the bright occulted
star.)  To account for some uncertainty in the exact
180\degs-symmetric point, we injected the simulated binary images at
each location in a 3$\times$3 pixel box centered on our best guess of
the 180\degs-symmetric location.  This also served to sample different
realizations of the noise being added to the simulated binary images.
We found that using a larger box size (e.g., 5$\times$5) did not
significantly change the results.

We then applied our PSF-fitting routine in an identical manner to the
simulated binary images as to the science data, with one exception.
Because the single star used to generate the simulated binary images
is actually taken from an $F814W$ image, we used the appropriate
$F814W$ TinyTim models in the simulations.  By comparing the input
separations, position angles (PAs, measured in degrees east from
north), and flux ratios of the simulated binaries to the fitted
values, we determined the random error and any significant systematic
offsets inherent in PSF-fitting routine.

We also investigated the effects of telescope defocus (e.g., due to
breathing) and jitter on our PSF-fitting.  In our analysis of
\HST/WFPC2 images of \twomassbin\ \citep{2008arXiv0807.0238L}, we
found that allowing these as free parameters in the PSF significantly
improved the residuals, producing slightly improved astrometric
precision.  For the 2002 observations of \hdbin, the best-fit defocus
and jitter of the science data reached unrealistic values
($>20$~\micron; $>20$~mas) when allowed as a free parameter, thus we
fixed telescope defocus and jitter to zero for both the science data
and the simulations of the 2002 epoch.  For the 2005 epoch, we found a
degeneracy between telescope defocus and the measured binary
separation, in the sense that tighter binaries could be equally well
fit if the amplitude of the defocus was allowed to be rather large
($+12$~\micron).  This may be intuitively understood since defocus
essentially increases the extent of the PSF, so for a given binary
footprint in a science image tighter separations can only be fit by
increasing the defocus.  The effect on the astrometry is indeed small
(0.05--0.08~pix, 1--2~mas) but significant.  Through the use of our
Monte Carlo simulations, in which we know the true separation of our
simulated binary images, we were able to break the degeneracy:
allowing defocus as a free parameter artificially decreased the
measured separation.  Only values of the defocus larger than are
typically observed ($\pm$10~\micron) produced this degeneracy.
Therefore we fixed defocus and jitter to zero for both the science
images and simulations of the 2005 epoch.

First, we consider the results for the 2002 epoch, where the science
data come from four different bandpasses: $F850LP$, $FR914M$
(8626~\AA), $FR914M$ (9402~\AA), and $FR914M$ (10248~\AA).  The
separation of \hdbin\ at this epoch is $\sim$2$\times$ larger than at
the 2005 epoch, but the $S/N$ is lower (see Figure~\ref{fig:images}),
so it is not clear that the astrometric precision should be better.
We found that the scatter among measurements taken in different
bandpasses was consistent with the random error predicted by the
simulations.  In Table~\ref{tbl:hst}, we quote the individual
measurements taken in different bandpasses with their respective
offsets applied and with uncertainties given from the Monte Carlo
simulations.  The systematic offsets were small compared to the
uncertainties (0.1--1.2$\sigma$).  To understand how to combine these
measurements, we investigated the nature of the uncertainties through
simulations where the $S/N$ was varied over the equivalent of 0.0--7.5
mag of noise degradation.  Since the $S/N$ of the single star used to
generate the simulated binaries is very high, this allows us to see
what the error would be if the $S/N$ were much higher or lower than
the science data.  If the astrometric uncertainties were truly
independent with respect to our PSF-fitting routine, they should
improve linearly with $S/N$ \citep{1983PASP...95..163K}.  We found
that at low $S/N$ the error in the 2002 epoch astrometry improved
slightly less than linearly, which implies a significant systematic
component to the error that cannot be reduced by averaging over
multiple measurements.  At high $S/N$, the error was constant over a
wide range of $S/N$, implying a systematic noise floor.  The $S/N$ of
the science data is near the boundary between these two error regimes,
thus the uncertainty of the science data is dominated by the
systematic component, and we cannot combine the individual
measurements in different bandpasses assuming they are independent.
Therefore, for the 2002 epoch, we use the single bandpass with the
smallest error in both separation and PA, $FR914M$ (9402~\AA), which
is also the bandpass with the cleanest background subtraction, highest
$S/N$, and lowest $\chi^2$ from PSF-fitting.

Given the higher $S/N$ of the 2005 data, our simulations show that our
astrometric uncertainty is better than for the 2002 epoch, even though
the binary is much tighter in the 2005 data.  All of the systematic
offsets predicted by our simulations were smaller than the predicted
random errors.  In Table~\ref{tbl:hst}, we quote the individual
measurements after applying all systematic offsets and give the
uncertainties determined from our simulations.  To understand how to
combine these measurements, we investigated the nature of the errors
through simulations where the $S/N$ was varied as described above for
the 2002 epoch.  We found that for all three 2005 measurements, the
science data fall almost exactly between two regimes: (1) at high
$S/N$ the errors improve slightly less than linearly implying both
random and systematic errors are significant; (2) at low $S/N$ the
errors improve linearly implying this regime is dominated by random
noise.  If our science data were in the noise-dominated regime, we
could hope to reduce the errors by $\sqrt{3}$ by averaging the three
measurements and adopting the standard error on the mean.  However,
since the data are in the high $S/N$ regime, no further reduction of
the uncertainty is possible due to a significant contribution from
non-independent systematic errors.  Therefore, we use the mean of the
three 2005 \HST\ measurements, with the typical (median) random error
predicted by the Monte Carlo simulations as its uncertainty.

\section{Monte Carlo Simulations of \hdbin\ in Keck NGS AO Images \label{app:keck}}

For each epoch of Keck observations, we conducted simulations to
determine the systematic and random errors inherent to our PSF-fitting
routine.  We created simulated binary images using the best available
empirical Keck NGS AO PSF.  The separation, PA, and flux ratio of each
simulated binary were randomly drawn to be within 0.3 pixels, 3\degs,
and 0.2 mag, respectively, of the measured values at that epoch.  We
used bilinear interpolation with cubic convolution to create a shifted
but otherwise identical PSF image.  The original and shifted PSFs were
each scaled to match the typical peak fluxes of \hdbin\ at the given
epoch.  Because the empirical PSF image used to construct the
simulated binary images was always of a higher $S/N$ than the science
data, random noise was added to the simulated binary images assuming
Poisson statistics for the science data and infinite $S/N$ for the
empirical input PSF.

The empirical input PSFs for the 2007 January epoch simulations were
images of the primary star \hdprim\ taken in $K_{cont}$-band directly
after obtaining $K$-band images of \hdbin.  The Strehl ratio and FWHM
of the images of \hdprim\ were 0.10$\pm$0.02 and 65.2$\pm$4.1 mas,
respectively.  In February 2007, the next generation wavefront
controller (NGWFC) was installed on Keck~II
\citep{2006SPIE.6272E...7W}.  Thus, for the remaining $K_S$- and
$K_{cont}$-band epochs, we used the NGWFC bright star $K$-band PSFs
($R=7.5$, 12.6, 13.6 mag) available on the Keck NGS AO
webpage\footnote{http://www2.keck.hawaii.edu/optics/ngsao/}.  We
measured the Strehl ratio and FWHM of this set of three PSFs to be
0.51$\pm$0.02 and 46.8$\pm$0.9 mas.  There is not a stark drop in
Strehl or FWHM with NGS brightness because the Keck NGS AO system
delivers similar on-axis image quality when the NGS is brighter than
$R$~$\approx$~13~mag.  This PSF stability is also what enables our use
of non-contemporaneous PSFs, which is further justified by the fact
that the systematic offsets derived from these empirical PSFs improves
the orbit fit (see \S \ref{sec:keck}).  Finally, we used Keck NGS
images of the bright star Gl~569A \citep[$R=9.4$
mag;][]{2003AJ....125..984M} taken in $H_{cont}$-band
($\lambda_c$~=~1.580~\micron, $\Delta\lambda$~=~0.023~\micron) on 2008
January 16 as the empirical input PSF for the 2007 July epoch
simulations.  The Strehl ratio and FWHM of these $H$-band PSFs were
0.29$\pm$0.04 and 37.9$\pm$0.6 mas, respectively.  In summary, while
the FWHM of the single PSFs we use in our simulations are comparable
to the FWHM of the science images, the Strehl ratios are consistently
somewhat worse than the science data (Table~\ref{tbl:keck}).
Therefore, we expect that the uncertainties from our Monte Carlo
simulations of the Keck astrometry may be slightly overestimated (this
is consistent with our best fit orbit, which has reduced $\chi^2<1$).

Background light due to the primary star \hdprim\ was added to the
simulated binary images in order to accurately replicate our science
images.  To accomplish this, we utilized the 60\degs-symmetric nature
of the hexagonal Keck PSF.  We extracted subregions from each dithered
image at 60\degs-symmetric locations relative to the location of
\hdprim.  Though the central pixels of \hdprim\ are saturated in these
images, its location can be deduced (typically to better than
$\approx$1 pixel) from the intersection of the 6 diffraction spikes.
The extracted subregions were appropriately rotated to match the
background at the location of \hdbin.  This yielded a number of
independent images of the background up to five times the number of
dithered images at a given epoch.  Some of the symmetric locations
fell off the array, but there was always at least one symmetric
location available for each dithered image.  For each epoch, 10$^3$
simulated science images were constructed by random pairing of the
extracted background images with the simulated binary images described
above.

The hexagonal Keck Airy pattern is clearly visible in all science
images, so care was taken to appropriately rotate the empirical input
PSFs to match the rotation in the science images.  This is important
because the overlap of the first hexagonal Airy ring of one component
with the core of the other is likely one of the major sources of
systematic error in our PSF-fitting technique.  The image rotator of
the Keck AO system changes to keep the PA of the sky fixed with
respect to NIRC2 during a set of observations.  Thus, the telescope
optics and resulting Airy pattern rotate with respect to NIRC2, and
these angles were measured directly from the header of each science
image.  For instance, the 2007 March 25 observations were conducted
over $\approx$8 minutes near transit, when the rotator is changing
fastest, so the PSF rotated by $\approx$7\degs\ with respect to the
sky during that time.

\clearpage

\clearpage
\begin{deluxetable}{lcccccc}
\tabletypesize{\footnotesize}
\rotate
\tablewidth{0pt}
\tablecaption{
\HST/ACS-HRC Coronagraph (1$\farcs$8 Spot) Observations \label{tbl:hst}}
\tablehead{
\colhead{Date/Start Time (UT)} &
\colhead{$t_{exp}$ (s)} &
\colhead{$\rho$ (mas)} &
\colhead{PA (\degs)} &
\colhead{$\Delta{f}$ (mag)} &
\colhead{Filter} &
\colhead{Note}}
\startdata
 2002 Sep 6/05:15 &  30 & 97.8$\pm$1.7 & 307.0$\pm$1.4 & 0.47$\pm$0.05 & $F850LP$             & 1 \\
 2002 Sep 6/04:51 & 200 & 99.3$\pm$1.1 & 307.9$\pm$1.1 & 0.72$\pm$0.05 & $FR914M$ (8626~\AA)  & 2 \\
 2002 Sep 6/04:56 & 200 & 94.6$\pm$1.1 & 306.9$\pm$1.0 & 0.41$\pm$0.04 & $FR914M$ (9402~\AA)  & 3 \\
 2002 Sep 6/05:01 & 200 & 96.1$\pm$1.7 & 308.9$\pm$1.6 & 0.38$\pm$0.08 & $FR914M$ (10248~\AA) & 4 \\
\tableline
 \textbf{2002 Sep 6} &  & \textbf{94.6$\pm$1.1} & \textbf{306.9$\pm$1.0} & \nodata  &  &  \\
\tableline
2005 Feb 23/15:39 & 300 & 56.8$\pm$0.6 & 144.9$\pm$0.5 & 0.19$\pm$0.03 & $F850LP$ & 1 \\
2005 Feb 23/17:15 & 300 & 56.4$\pm$0.9 & 146.4$\pm$0.6 & 0.24$\pm$0.06 & $F850LP$ & 1 \\
2005 Feb 23/18:46 & 300 & 57.3$\pm$0.6 & 148.6$\pm$0.6 & 0.30$\pm$0.05 & $F850LP$ & 1 \\
\tableline
 \textbf{2005 Feb 23} & & \textbf{56.8$\pm$0.6} & \textbf{146.6$\pm$0.6} & \textbf{0.24$\pm$0.05} & \textbf{$F850LP$}  &  \\
\enddata

\tablecomments{Source of image used for background subtraction:
  (1)---\hdprim, $F850LP$, 2005 Feb 23 UT;
  (2)---$\alpha$ Cen A, $FR914M$ (8495~\AA), 2004 Aug 7 UT;
  (3)---$\alpha$ Cen A, $FR914M$ (9282~\AA), 2004 Aug 7 UT;
  (4)---$\alpha$ Cen A, $FR914M$ (10500~\AA), 2003 Sep 12 UT.}

\end{deluxetable}

\begin{deluxetable}{lcccccccc}
\tabletypesize{\footnotesize}
\rotate
\tablewidth{0pt}
\tablecaption{Keck NGS AO Observations \label{tbl:keck}}
\tablehead{
\colhead{Date/Start Time (UT)} &
\colhead{Filter\tablenotemark{a}} &
\colhead{$N \times t_{exp}$ (s)\tablenotemark{b}} &
\colhead{Airmass} &
\colhead{FWHM (mas)\tablenotemark{c}} &
\colhead{Strehl ratio\tablenotemark{c}} &
\colhead{$\rho$ (mas)\tablenotemark{d}} &
\colhead{P.A. (\degs)\tablenotemark{d}} &
\colhead{$\Delta{f}$ (mag)}}
\startdata
2007 Jan 26/14:03 & $K$    & 10$\times$36 & 1.28 & 57.0$\pm$4.0 & 0.26$\pm$0.11 & 111.7$\pm$0.8  & 132.6 $\pm$0.4  & 0.25 $\pm$0.03  \\
2007 Mar 25/12:41 & \Kcont &  6$\times$30 & 1.01 & 47.8$\pm$0.1 & 0.64$\pm$0.02 & 109.1$\pm$0.4  & 132.28$\pm$0.13 & 0.19 $\pm$0.02  \\
2007 Jul 25/05:41 & $H$    &  9$\times$30 & 1.01 & 35.7$\pm$0.6 & 0.42$\pm$0.09 &  97.9$\pm$0.3  & 130.73$\pm$0.17 & 0.19 $\pm$0.05  \\
2008 Jan 15/16:37 & \Ks    & 12$\times$30 & 1.03 & 48.0$\pm$0.7 & 0.58$\pm$0.07 &  72.2$\pm$0.2  & 127.6 $\pm$0.3  & 0.15 $\pm$0.03  \\
2008 Mar 29/13:38 & $K$    & 14$\times$30 & 1.03 & 49.1$\pm$0.3 & 0.65$\pm$0.02 &  57.4$\pm$0.2  & 124.7 $\pm$0.4  & 0.197$\pm$0.003 \\
2008 Mar 29/14:09 & $J$    &  8$\times$60 & 1.07 & 32.4$\pm$0.8 & 0.28$\pm$0.02 &  57.3$\pm$0.6  & 124.6 $\pm$0.6  & 0.305$\pm$0.014 \\
2008 Mar 29/15:46 & $H$    &  6$\times$30 & 1.36 & 41.8$\pm$1.1 & 0.35$\pm$0.03 &  58.4$\pm$0.6  & 124.1 $\pm$0.7  & 0.29 $\pm$0.02  \\
2008 Apr 27/14:32 & \Ks    &  3$\times$30 & 1.61 & 48.5$\pm$0.5 & 0.63$\pm$0.03 &  51.7$\pm$0.3  & 123.9 $\pm$0.5  & 0.199$\pm$0.005 \\
\enddata

\tablenotetext{a}{All photometry on the MKO system.}

\tablenotetext{b}{ $N$ is the number of dithered images, each of
  exposure time $t_{exp}$, taken at that epoch. }

\tablenotetext{c}{Computed as described in the text using a
  0$\farcs$75 aperture, except for the 2007 Jan 26 epoch which
  required a smaller aperture (0$\farcs$5) because poor image quality
  led to increased flux contamination from \hdprim.  The quoted value
  and its error correspond to the mean and RMS of the set of dithered
  images.}

\tablenotetext{d}{The tabulated errors are computed by adding in
  quadrature the uncertainty in the NIRC2 pixel scale and orientation
  and the uncertainty that is predicted for each epoch from the Monte
  Carlo simulations described in the text.  We used a weighted average
  of the astrometric calibration from \citet{2006ApJ...649..389P},
  with a pixel scale of 9.963$\pm$0.011 mas/pixel and an orientation
  for the detector's $+y$-axis of $-$0.13$\pm$0.07 east of north.}
\end{deluxetable}

\begin{deluxetable}{lcccc}
\tabletypesize{\small}
\tablewidth{0pt}
\tablecaption{Near-infrared MKO Photometry of HD~130948ABC \label{tbl:phot}}
\tablehead{
\colhead{Property} &
\colhead{HD~130948A} &
\colhead{HD~130948B} &
\colhead{HD~130948C} &
\colhead{Reference}}
\startdata
$J$ (mag)  & 4.797$\pm$0.022 &  13.81$\pm$0.06  &  14.12$\pm$0.06  &  1,2    \\
$H$ (mag)  & 4.515$\pm$0.022 &  13.04$\pm$0.10  &  13.33$\pm$0.11  &  1,2    \\
$K$ (mag)  & 4.458$\pm$0.020 &  12.26$\pm$0.03  &  12.46$\pm$0.03  &  1,2    \\
\enddata

\tablerefs{(1) This work; (2) \citet{2mass-2}.}
\end{deluxetable}

\begin{deluxetable}{lcccc}
\tabletypesize{\footnotesize}
\rotate
\tablewidth{0pt}
\tablecaption{Derived Orbital Parameters for \hdbin \label{tbl:orbit}}
\tablehead{
\colhead{}   &
\multicolumn{3}{c}{MCMC}      &
\colhead{\orbit\tablenotemark{\dag}} \\
\cline{2-4}
\colhead{Parameter}   &
\colhead{Median}      &
\colhead{68.3\% c.l.} &
\colhead{95.5\% c.l.} &
\colhead{} }

\startdata
Semimajor axis $a$ (mas)                                         &  121   &      $-$6, 6      &  $-$10, 14        &    121$\pm$7      \\
Orbital period $P$ (yr)                                          &  9.9   &    $-$0.6, 0.7    &  $-$1.1, 1.6      &    9.9$\pm$0.8    \\
Eccentricity $e$                                                 & 0.167  &  $-$0.015, 0.020  &  $-$0.03, 0.05    &  0.163$\pm$0.019  \\
Inclination\tablenotemark{a} $i$ (\degs)                         &  95.7  &    $-$0.2, 0.3    &  $-$0.5, 0.5      &   95.8$\pm$0.3    \\
Time of periastron passage $T_0-2454664.0$\tablenotemark{b} (JD) &   0    &    $-$110, 110    &  $-$200, 200      &     14$\pm$130    \\
PA of the ascending node $\Omega$ (\degs)                        & 133.15 &   $-$0.16, 0.15   &  $-$0.3, 0.3      & 133.14$\pm$0.15   \\
Argument of periastron $\omega$ (\degs)                          &   71   &     $-$14, 15     &  $-$30, 30        &     73$\pm$18     \\
Total mass (\Msun): fitted\tablenotemark{c}                      & 0.1085 & $-$0.0017, 0.0019 &  $-$0.003, 0.004  & 0.1083$\pm$0.0019 \\
Total mass (\Msun): final\tablenotemark{d}                       & 0.109  &  $-$0.002, 0.002  &  $-$0.004, 0.005  &  0.108$\pm$0.002  \\
Reduced $\chi^2$                                                 &  1.14  &       \nodata     &       \nodata     &        1.14       \\
\enddata

\tablenotetext{\dag}{The orbital parameters determined by \orbit\ are
  listed along with their linearized 1$\sigma$ errors.}

\tablenotetext{a}{By convention, $i$~$>$~90\degs\ denotes that the sky
  PA is decreasing over time (clockwise motion), rather than
  increasing (counterclockwise).}

\tablenotetext{b}{2008 Jul 16 12:00:00.0 UT} 

\tablenotetext{c}{The ``fitted'' total mass represents the direct
  results from fitting the observed orbital motion of the two
  components.  For the linearized \orbit\ error, the covariance
  between $P$ and $a$ is taken into account.}

\tablenotetext{d}{The ``final'' total mass includes the additional
  1.3\% error in the mass from the \Hipparcos\ parallax of \hdprim.
  This final mass distribution is essentially Gaussian.}

\end{deluxetable}

\begin{deluxetable}{lccccc}
\tabletypesize{\footnotesize}
\tablewidth{0pt}
\tablecaption{Age Estimates for \hdprim \label{tbl:age}}
\tablehead{
\colhead{}   &
\multicolumn{4}{c}{Age (Gyr)} &
\colhead{}  \\
\cline{2-5}
\colhead{Age indicator}   &
\colhead{Estimate}      &
\colhead{68.3\% c.l.} &
\colhead{95.5\% c.l.} &
\colhead{Precision (1$\sigma$)} &
\colhead{Ref.} }
\startdata
Stellar rotation (``gyrochronology'')             &    0.79    &   0.64--1.01    &   0.53--1.32  &  $\approx$25\%       &     1     \\
                                                  &    0.65    &   0.55--0.78    &   0.47--0.93  &  $\approx$20\%       &     2     \\
Chromospheric activity (\ion{Ca}{2}~HK emission)  &    0.5     &    0.2--0.8     &    \nodata    &  $\approx$60\%       &     1     \\
Stellar isochrones                                &    0.72    &    \nodata      &   0.32--2.48  &  $\approx$2$\times$  &     3     \\
X-ray activity\tablenotemark{\dag}                &   \nodata  &    0.1--0.6     &    \nodata    &  \nodata             &  1,4,5,6  \\
Lithium depletion\tablenotemark{\dag}             &   \nodata  &    0.1--0.6     &    \nodata    &  \nodata             &     7     \\
\enddata
\tablenotetext{\dag}{These indicators do not give quantifiable age
  estimates or corresponding uncertainties.  They do show that the age
  of \hdprim\ is generally consistent with stars in the Hyades and
  Pleiades, thus we adopt these clusters' ages as the 1$\sigma$
  range.}  \tablerefs{(1) \citet{mam08-ages}; (2)
  \citet{2007ApJ...669.1167B}; (3) \citet{2006astro.ph..7235T}; (4)
  \citet{1995ApJ...448..683S}; (5) \citet{1998PASP..110.1259G}; (6)
  \citet{2001A&A...377..538S}; (7) \citet{1993AJ....106.1080S,
    1993AJ....106.1059S, 1993AJ....105.2299S}.}
\end{deluxetable}

\begin{deluxetable}{lccc}
\tablewidth{0pt}
\tablecaption{Measured Properties of \hdbin \label{tbl:meas}}
\tablehead{
\colhead{Property}   &
\colhead{HD~130948B} &
\colhead{HD~130948C} &
\colhead{Note}       }
\startdata
$M_{tot}$ (\Msun)           & \multicolumn{2}{c}{   0.109$\pm$0.002  }  &  1  \\
$d$ (pc)                   & \multicolumn{2}{c}{   18.18$\pm$0.08   }  &  2  \\
Spectral Type              &      L4$\pm$1     &      L4$\pm$1         &  3  \\
BC$_K$ (mag)               &    3.33$\pm$0.13  &    3.34$\pm$0.13      & 1,4 \\
$\Delta{{\rm BC}_K}$ (mag) &  \multicolumn{2}{c}{   0.00$\pm$0.02   }  & 1,4  \\
$J-H$ (mag)                &    0.77$\pm$0.12  &    0.79$\pm$0.13      &  1  \\
$H-K$ (mag)                &    0.78$\pm$0.10  &    0.87$\pm$0.11      &  1  \\
$J-K$ (mag)                &    1.55$\pm$0.07  &    1.66$\pm$0.07      &  1  \\
$M_J$ (mag)                &   12.51$\pm$0.06  &   12.82$\pm$0.06      &  1  \\
$M_H$ (mag)                &   11.74$\pm$0.10  &   12.03$\pm$0.11      &  1  \\
$M_K$ (mag)                &   10.96$\pm$0.03  &   11.16$\pm$0.03      &  1  \\
$\Delta{K}$ (mag)          &  \multicolumn{2}{c}{  0.197$\pm$0.008  }  &  1  \\
$\log$(\Lbol/\Lsun)        & $-$3.82$\pm$0.05  & $-$3.90$\pm$0.05      &  1  \\
$\Delta\log$(\Lbol)        &  \multicolumn{2}{c}{  0.079$\pm$0.008  }  &  1  \\
\enddata
\tablerefs{(1) This work; (2) \citet{2007hnrr.book.....V}; (3)
  \citet{2002ApJ...567L..59G}; (4) \citet{gol04}.}
\end{deluxetable}

\begin{deluxetable}{lccc}
\tabletypesize{\scriptsize}
\tablewidth{0pt}
\tablecaption{Evolutionary Model Inferred Properties of \hdbin \label{tbl:model}}
\tablehead{
\colhead{Property}    &
\colhead{Median}      &
\colhead{68.3\% c.l.} &
\colhead{95.5\% c.l.} }
\startdata
                           \multicolumn{4}{c}{\bf Tucson Models \citep{1997ApJ...491..856B}} \\
                           \cline{1-4}
                           \multicolumn{4}{c}{System} \\
                           \cline{1-4}
Age (Gyr)                     & 0.41    & $-   0.03, 0.04   $ & $-   0.06, 0.08   $ \\
$q$ ($M_{\rm C}/M_{\rm B}$)   & 0.962   & $-  0.003, 0.003  $ & $-  0.009, 0.009  $ \\
$\Delta$\Teff\ (K)            & 90      & $-      8, 7      $ & $-     22, 22     $ \\
Li$_{\rm C}$/Li$_{\rm B}$\tablenotemark{\dag}     & 1.049   & $-  0.014, 0.011  $ & $-   0.02, 0.03   $ \\
                           \cline{1-4}
                           \multicolumn{4}{c}{Component B} \\
                           \cline{1-4}
$M_{\rm B}$ (\Msun)           & 0.0554  & $- 0.0013, 0.0012 $ & $-  0.002, 0.002  $ \\
$T_{\rm eff,B}$ (K)           & 2040    & $-     50, 50     $ & $-    110, 110    $ \\
$\log(g_{\rm B})$ (cgs)       & 5.196   & $-  0.017, 0.017  $ & $-   0.03, 0.03   $ \\
$R_{\rm B}$ (\Rsun)           & 0.0983  & $- 0.0011, 0.0011 $ & $-  0.002, 0.002  $ \\
Li$_{\rm B}$/Li$_0$           & 0.91    & $-   0.03, 0.03   $ & $-   0.07, 0.05   $ \\
                           \cline{1-4}
                           \multicolumn{4}{c}{Component C} \\
                           \cline{1-4}
$M_{\rm C}$ (\Msun)           & 0.0532  & $- 0.0011, 0.0012 $ & $-  0.002, 0.002  $ \\
$T_{\rm eff,C}$ (K)           & 1950    & $-     50, 50     $ & $-    110, 110    $ \\
$\log(g_{\rm C})$ (cgs)       & 5.179   & $-  0.017, 0.017  $ & $-   0.03, 0.03   $ \\
$R_{\rm C}$ (\Rsun)           & 0.0983  & $- 0.0010, 0.0010 $ & $-  0.002, 0.002  $ \\
Li$_{\rm C}$/Li$_0$           & 0.96    & $-   0.02, 0.02   $ & $-   0.05, 0.03   $ \\
                           \cline{1-4}
                           \multicolumn{4}{c}{} \\
                           \multicolumn{4}{c}{\bf Lyon Models \citep[DUSTY;][]{2000ApJ...542..464C}} \\
                           \cline{1-4}
                           \multicolumn{4}{c}{System} \\
                           \cline{1-4}
Age (Gyr)                     & 0.45    & $-   0.04, 0.05   $ & $-   0.08, 0.10   $ \\
$q$ ($M_{\rm C}/M_{\rm B}$)   & 0.948   & $-  0.005, 0.005  $ & $-  0.013, 0.012  $ \\
$\Delta$\Teff\ (K)            & 85      & $-      7, 7      $ & $-     21, 21     $ \\
Li$_{\rm C}$/Li$_{\rm B}$\tablenotemark{\dag}     & 1.6     & $-    0.3, 1.0    $ & $-    0.5, 3.5    $ \\
                           \cline{1-4}
                           \multicolumn{4}{c}{Component B} \\
                           \cline{1-4}
$M_{\rm B}$ (\Msun)           & 0.0558  & $- 0.0012, 0.0012 $ & $-  0.002, 0.002  $ \\
$T_{\rm eff,B}$ (K)           & 1990    & $-     50, 50     $ & $-    100, 90     $ \\
$\log(g_{\rm B})$ (cgs)       & 5.143   & $-  0.019, 0.019  $ & $-   0.04, 0.04   $ \\
$R_{\rm B}$ (\Rsun)           & 0.1048  & $- 0.0016, 0.0017 $ & $-  0.003, 0.004  $ \\
Li$_{\rm B}$/Li$_0$           & 0.50    & $-   0.23, 0.18   $ & $-    0.4, 0.3    $ \\
                           \cline{1-4}
                           \multicolumn{4}{c}{Component C} \\
                           \cline{1-4}
$M_{\rm C}$ (\Msun)           & 0.0528  & $- 0.0012, 0.0012 $ & $-  0.002, 0.002  $ \\
$T_{\rm eff,C}$ (K)           & 1900    & $-     50, 50     $ & $-    100, 90     $ \\
$\log(g_{\rm C})$ (cgs)       & 5.122   & $-  0.019, 0.019  $ & $-   0.04, 0.04   $ \\
$R_{\rm C}$ (\Rsun)           & 0.1045  & $- 0.0015, 0.0016 $ & $-  0.003, 0.003  $ \\
Li$_{\rm C}$/Li$_0$           & 0.83    & $-   0.13, 0.08   $ & $-   0.31, 0.12   $ \\
\enddata
\tablenotetext{\dag}{Ratio of the lithium abundance of the two
  components.}
\end{deluxetable}

\end{document}